\documentclass[11pt]{article}
\usepackage{jheppub}
\usepackage{bbm}
\usepackage[utf8]{inputenc}
\usepackage{slashed} 
\usepackage{ytableau}

\makeatletter

\usepackage{tikz}
\usetikzlibrary{intersections, calc,positioning,decorations.pathreplacing,decorations.pathmorphing}

\def\D{{\Delta}}
\def\bomega{{\boldsymbol \omega}}

\def\CR{{\cal R}}
\def\CM{{\cal M}}
\def\CN{{\cal N}}

\def\BC{\mathbb{C}}

\def\BS{\mathbb{S}}
\def\BZ{\mathbb{Z}}

\renewcommand{\[}{\begin{equation}}
\renewcommand{\]}{\end{equation}}
\numberwithin{equation}{section}

\title{Supersymmetric R{\'e}nyi Entropy and Defect Operators}

\author[a]{Tatsuma Nishioka}
\author[b]{and Itamar Yaakov}

\affiliation[a]{Department of Physics, Faculty of Science, The University of Tokyo, \\Bunkyo-ku, Tokyo 113-0033, Japan}
\affiliation[b]{Kavli IPMU (WPI), UTIAS, The University of Tokyo, Kashiwa, Chiba 277-8583, Japan}

\emailAdd{nishioka@hep-th.phys.s.u-tokyo.ac.jp}
\emailAdd{itamar.yaakov@ipmu.jp}

\abstract{We describe the defect operator interpretation of the
supersymmetric R{\'e}nyi entropies of superconformal
field theories in three, four and five dimensions. The operators involved
are supersymmetric codimension-two defects in an auxiliary $\mathbb{Z}_{n}$
gauge theory coupled to $n$ copies of the SCFT. We compute the exact
expectation values of such operators using localization, and compare
the results to the supersymmetric R{\'e}nyi entropy.
The agreement between the two implies a relationship between the partition
function on a squashed sphere and the one on a round sphere in the
presence of defects.}

\preprint{UT-16-35, IPMU16-0191}

\begin{document}
\maketitle

\section{Introduction}

The non-local nature of quantum entanglement is one of the sharpest
characteristics by which quantum physics differentiates itself from
classical physics. Entanglement occupies a central position in quantum
information theory and, increasingly, in various branches of theoretical
physics such as condensed matter and high energy physics. Entangled
states are ubiquitous and of particular interest in many-body quantum
systems. R{\'e}nyi entropy is a refined measure of
the entanglement a given state possesses when the Hilbert space is
split into states supported on a spatial region $\Sigma$ and those
supported on its complement. In a local quantum field theory, employing
the replica trick \cite{Calabrese:2004eu}, the $n^{th}$ R{\'e}nyi
entropy amounts to 
\begin{align}
S_{n}\equiv\frac{1}{1-n}\log\left|\frac{Z_{n}}{\left(Z_{1}\right){}^{n}}\right|\ .\label{Renyi_Definition}
\end{align}
The partition function $Z_{n}$ is defined on an $n$-fold cover $\CM_{n}$
branched over the entangling surface $\partial\Sigma$ of a (Euclidean)
manifold on which the theory is placed. The absolute value taken in
the definition (\ref{Renyi_Definition}) has no effect in unitary
theories, but is necessary to incorporate the case of a complex partition
function which we will deal with when supersymmetry is implemented
on a curved space.

It follows from (\ref{Renyi_Definition}) that knowing the partition
function $Z_{n}$ is more or less equivalent to calculating R{\'e}nyi
entropies, and there are a few situations where the exact values are
known (see e.g.\ \cite{Calabrese:2004eu,Casini:2010kt,Klebanov:2011uf}).
A common practice in handling the conical singularity around $\partial\Sigma$,
present in the calculation of $Z_{n}$ for $n>1$, is to smooth out
the tip by introducing a regulator, calculate the partition function
on the smoothed space, and take the singular limit \cite{Fursaev:1995ef}.
This approach is highly advantageous as it reformulates the problem
as a calculation in quantum field theory on a curved space. 

Another complementary approach is to represent the partition function
$Z_{n}$ as a product of correlation functions of twist operators
that create the proper monodromies around $\partial\Sigma$ \cite{Calabrese:2004eu,Hung:2011nu}.
Twist operators are codimension-two (non-local in $d>2$ dimensions)
objects that specify the boundary conditions on the entangling surface,
and the twisting is done for an $n$-fold copy of the original theory.
In what follows, we will illustrate the interplay between the two
approaches in a particular situation where the exact calculation of
$Z_{n}$ is possible: the supersymmetric R{\'e}nyi
entropy computed using localization \cite{Nishioka:2013haa}. We will restrict ourselves to
a spherical entangling surface $\partial\Sigma=\BS^{d-2}$ in $d$
dimensions, work in the vacuum state of the SCFT, and examine supersymmetric
gauge theories of type $3d$ $\mathcal{N}=2$, $4d$ $\mathcal{N}=2$,
and $5d$ $\mathcal{N}=1$. The motivation for doing so is twofold.
First, it is interesting to compare the two different looking localization
calculations and find out how they match. Second, the microscopic
definition of the defect operators,\footnote{Twist operators are a subclass of defect operators. These terms will be used interchangeably in this paper.} presented in Section \ref{sec:Defects_from_abelian_duality},
could be useful when examining dualities in which the SCFT participates.
These could be dualities between different Lagrangian field theories,
or holographic dualities between an SCFT and string theory on an appropriate
background.

\subsection{Supersymmetric R{\'e}nyi entropy and defects}

The vacuum R{\'e}nyi entropy is a non-local observable
which can be defined for any $d$-dimensional quantum field theory
whose Hilbert space can be factorized in a local manner. For a conformal
field theory, and for integer R{\'e}nyi parameter
$n$, this observable is equivalent to two different objects, each
of which can be defined by a suitable Euclidean path integral:
\begin{enumerate}
\item[I.] The partition function on an $n$-fold multiple covering of the $d$-sphere
$\mathbb{S}^{d}$, branched along $\partial\Sigma$, with appropriate
boundary conditions at the branching loci \cite{Casini:2011kv}.
\item[II.] The partition function of an $n$-fold copy of the theory - henceforth
referred to as the $n$-copy theory - on $\mathbb{S}^{d}$
with codimension-one defect operators acting between copies (c.f.
\cite{Bianchi:2015liz}).
\end{enumerate}
The equivalence between these two objects is tautological. The object
in II is simply a relabeling of the degrees of freedom, one for each
of the $n$ sheets of the branched sphere. The defect operators are
defined to reproduce the boundary conditions implied by the original
geometry. 

One may introduce a linear field redefinition, acting in the $n$-copy
theory, to diagonalize the action of the defects, introducing defects
which act on just one copy at a time. In the special case of a free
theory, the action written using the redefined fields does not couple
the copies. The computation in this case is equivalent to a third
object:
\begin{enumerate}
\item [III.]\setcounter{enumi}{3}The partition function of an $n$-fold copy
of the theory on $\mathbb{S}^{d}$ with codimension-two defect operators
acting on each copy - equivalently, the $k^{th}$ copy is coupled to
a background connection with holonomy $e^{2\pi i\frac{k}{n}}$ around
$\partial\Sigma$, with $k\in\left\{ 0,\ldots, n-1\right\} $.\footnote{The holonomy prescription could be different if the field in question
is a fermion, or carries additional global symmetry charges \cite{Casini:2009sr}.
We will make a specific choice later on.} This is really a product of the $n$ partition functions, since the
copies are decoupled. 
\end{enumerate}
We have codimension-two defects since the partition function over
free fields, which are subject to a monodromy when circling $\partial\Sigma$,
does not care about the position of $\Sigma$.

One may also alter the definition of the R{\'e}nyi
entropy to preserve additional symmetries of the theory. Specifically,
a subset of superconformal symmetry can be preserved by altering the
boundary conditions in the object I, II, or III (equivalently, changing the defects).
The resulting observable is called the supersymmetric R{\'e}nyi
entropy \cite{Nishioka:2013haa} (see also \cite{Huang:2014gca,Nishioka:2014mwa,Crossley:2014oea,Huang:2014pda,Hama:2014iea,Alday:2014fsa,Giveon:2015cgs,Mori:2015bro,Nian:2015xky} for further developments).\footnote{Some of the $3d$ theories we consider are not superconformal. However,
deformation invariance of the partition function is enough to guarantee
that the result can then be interpreted as the supersymmetric R{\'e}nyi
entropy of the SCFT to which the original theory flows \cite{Nishioka:2013haa}.} 
It has been observed that the result in this case is equivalent to yet another object:
\begin{enumerate}
\item [IV.]\setcounter{enumi}{4}The partition function of a single copy
of the theory on a squashed $\mathbb{S}^{d}$, with a suitable supersymmetry
preserving action. The squashing parameter, which determines the
non-round metric, is related to the R{\'e}nyi parameter
$n$ in a simple way, such that $n=1$ corresponds to the round sphere.
\end{enumerate}
The simplest explanation for the last equivalence is as follows. 
Preserving supersymmetry means that the Euclidean action is invariant under some
supersymmetry transformation $\delta$. It is well known that the
path integral in this case is insensitive to $\delta$-exact deformations,
either of the action or in the form of $\delta$-exact insertions.\footnote{There are additional restrictions on the deformation. A $\delta$-exact deformation to the action must be annihilated by $\delta$ and
should have a positive semi-definite real part. Specifically, any
deformation must be such that it does not alter the convergence properties
of the the path integral.}
The objects I and IV are related by such a deformation (c.f \cite{Closset:2013vra}). 

Despite their simplicity, there is something interesting to be said
about forms III and IV of the supersymmetric R{\'e}nyi
entropy. The calculation of the path integrals representing either
III or IV can be performed exactly using localization. This involves
splitting the fields of the theory into an interacting part - the
moduli - and a free part - the fluctuations. The latter can be put
into the form III. Since the two calculations look quite different,
and in some contexts have different interpretations, it is interesting
and potentially useful to determine exactly how they give the same
result. This will be our primary goal.

\subsection{\label{sub:Renyi-entropy-and}R{\'e}nyi entropy and discrete gauge theories}

The setup described above for calculating the R{\'e}nyi
entropy at integer R{\'e}nyi parameter can be alternatively
thought of as introducing a defect in a discrete gauge theory coupled
to the $n$-copy theory.\footnote{The authors are grateful to Daniel Jafferis for suggesting a discrete
gauge theory interpretation for our calculation.} For our purposes, it is sufficient to consider the gauge group $\mathbb{Z}_{n}$
acting by cyclic permutation on the copies, although the $n$-copy
theory is invariant under the full permutation group ${\mathfrak S}_{n}$.\footnote{The symmetry which shifts the copies by one is known as the replica $\BZ_n$ symmetry.}
If we choose to think of $\mathbb{Z}_{n}$ as a gauge symmetry, we
can reasonably treat the defects which implement the calculation of
the R{\'e}nyi entropy as codimension-two objects. Only operators charged 
under the $\mathbb{Z}_{n}$ symmetry can detect the position of the codimension-one 
defect. Gauging $\mathbb{Z}_{n}$ means that all such operators are projected out. 


We would now like to incorporate supersymmetry into the definition
of the $\mathbb{Z}_{n}$ gauge theory and the defect. A simple way
of doing so is to write down a version of the discrete gauge theory
which is realized by higher form abelian gauge fields \cite{Kapustin:2014gua}.
A BF type theory with one ordinary gauge field ($A$) and one $(d-2)$-form field ($B)$ works nicely. The reduction to $\mathbb{Z}_{n}$
gauge symmetry is implemented by using $B$ as a Lagrange multiplier.
The action for the gauge sector is
\begin{equation}
S_{\text{BF}}=\frac{in}{2\pi}\int F\wedge B\ ,\label{eq:topological_action}
\end{equation}
where $F$ is the field strength of $A$. $S_{\text{BF}}$ is invariant
under gauge transformations for both $A$ and $B$ as long as $n$
is an integer.

The codimension-two defect could be realized by taking a flat connection
with prescribed holonomy and coupling it to the free fields of the
$n$-copy theory. The way to do this in the BF theory is to insert
the operator
\begin{equation}
\exp\left(ik\oint_{\partial\Sigma}B\right)=\exp\left(ik\int B\wedge\left[\partial\Sigma\right]\right),\label{eq:Wilson_operator}
\end{equation}
where $\left[\partial\Sigma\right]$ is the Poincar{\'e} dual to the cycle
$\partial\Sigma$ representing the entangling surface. This operator is invariant
under the $B$ gauge transformations as long as $k$ is an integer.
Integrating out $B$ in the theory with the action \eqref{eq:topological_action},
and in the presence of this operator, induces a holonomy for $A$,
around $\partial\Sigma$ of strength $e^{2\pi i\frac{k}{n}}$. After
the linear field redefinition, $A$ can serve as the flat connection
for the $k^{th}$ copy for the R{\'e}nyi parameter $n$.
One may then extend the fields $A,B$ and the terms in the action
to their supersymmetric versions to create a SUSY-BF theory \cite{Brooks:1994nn}.\footnote{By SUSY-BF, we mean an untwisted supersymmetric version of BF theory
of the type considered in \cite{Brooks:1994nn}. We are not concerned,
in this paper, with the topological field theory aspects of supersymmetric
BF theories - which is the context in which they were originally introduced.
Specifically, we couple the SUSY-BF fields to the $n$-copy theory
and introduce localizing terms, both of which can spoil the topological
properties. }

One possible obstruction to using ordinary gauge fields is that the
original theory may have its own, possibly non-abelian, gauge fields.
It may not be possible to couple the field $A$ to these non-abelian
gauge fields in a manner consistent with gauge invariance. Note that
the original description of the codimension-one defect already implies
that any gauge group of the $n$-copy theory is broken to a diagonal
subgroup, acting simultaneously and identically on the copy at either
side, at the defect. This type of the reduction of the gauge invariance
at the position of a defect is quite common (see e.g. \cite{Gaiotto:2008ak}).
Moreover, gauge fields often carry supersymmetric moduli which cannot
be coupled to the codimension-two defect in the way described above.
We will treat all of these problems in an ad hoc manner. 

The organization of the paper is as follows. In Section \ref{sec:Defects_from_abelian_duality}
we introduce a supersymmetric version of the $\mathbb{Z}_{n}$ gauge theory
based on higher form abelian gauge superfields. Component expressions
for the superfields and actions are readily available in the literature.
We then derive the abelian version of the codimension-two supersymmetric
defect operator as a classical supersymmetric configuration for a
vector supermultiplet on the round $d$-dimensional sphere. We use
this configuration to write down an ordinary supersymmetric Wilson
type codimension-two operator for the tensor multiplet, using the
supersymmetric version of \eqref{eq:Wilson_operator}, and perform localization
in the presence of this operator and the BF term. We then derive the
resulting effect on any matrix model to which the original vector
multiplet could potentially be coupled. In Section \ref{sec:Squashing_from_defects},
we show that the matrix model for the $n$-copy theory on the round
$\mathbb{S}^{d}$, after a specific modification by such abelian defects,
is equal to the matrix model on the squashed $\mathbb{\mathbb{S}}^{d}$,
including all classical and one-loop contributions. In the process,
we show how the moduli for the squashed sphere can be thought of as
$n$ sets of moduli for the round sphere, which are sewn together
by the defect. We end with a discussion of possible applications.

\section{\label{sec:Defects_from_abelian_duality}Supersymmetric defects from supersymmetric $\mathbb{Z}_{n}$ gauge theory}

In this section, we define the supersymmetric $\mathbb{Z}_{n}$ gauge
theory, and supersymmetric codimension-two defects, which we later
use to calculate the supersymmetric R{\'e}nyi entropy.
We then calculate the expectation values of these defects using localization.
We treat supersymmetric gauge theories of type $3d$ $\mathcal{N}=2$,
$4d$ $\mathcal{N}=2$, and $5d$ $\mathcal{N}=1$.

\subsection{Supersymmetric $\mathbb{Z}_{n}$ gauge theory}

We construct a supersymmetric $\mathbb{Z}_{n}$ gauge theory, and
a supersymmetric codimension-two defect, by introducing a pair of
dynamical supermultiplets: a vector multiplet $V$ and a $(d-2)$-form
multiplet $E$. The field strength for the latter sits in the familiar
linear multiplet, denoted by $G$. We introduce the supersymmetric
analogues of the terms used in the discrete gauge theory
\[
S_{\text{BF}}\quad \rightarrow\quad \frac{in}{2\pi}\int G\,V\ ,
\]
and 
\begin{equation}
\exp\left(ik\oint_{\partial\Sigma}B\right)\quad\rightarrow\quad \exp\left(ik\oint_{\partial\Sigma}E\right).\label{eq:supersymmetric_electric_operator}
\end{equation}
The terms on the right hand side are schematic superspace integrals, for which
we write component expressions later. 
The operator \eqref{eq:supersymmetric_electric_operator}
can also be written as
\[
\exp\left(\frac{ik}{2\pi}\int G\,V_{\text{defect}}\right)\ ,
\]
where $V_{\text{defect}}$ is a background configuration for a vector
multiplet invariant under a subset of the supersymmetries, whose component
expression will be worked out below, and the integral is over the
entire superspace. Integrating out $G$ results in a supersymmetric delta function setting 
\[
V=\frac{k}{n}\,V_{\text{defect}}\ .
\]
We call the coefficient
\[
q_{k}^{\text{vortex}}=\frac{k}{n} \ ,
\]
the vortex charge.

\subsection{\label{sub:3d_defect}The $3d$ $\mathcal{N}=2$ codimension-two defect}

In three dimensions, both $V$ and $E$ are ordinary vector multiplets.
The superfield $G$ is the field strength superfield which can be
used to write the Yang-Mills term in 3$d$.\footnote{This is a real linear superfield, which is sometimes denoted by $\Sigma$.}
The codimension-two defect is a vortex loop of the type examined,
for instance, in \cite{Moore:1989yh}. Its supersymmetric version
was analyzed in \cite{Kapustin:2012iw,Drukker:2012sr}. The calculation
of the supersymmetric R{\'e}nyi entropy in three
dimensions using defects was carried out in \cite{Nishioka:2013haa}.
We briefly review it below.

The localization calculation for an $\mathcal{N}=2$ theory on the
round $\BS^3$ reduces the path integral to an integration over
a single Lie algebra valued scalar: the constant mode 
\[
\sigma=-D\ ,
\]
of the \emph{real} fields $\sigma$ and $D$ appearing in the $\mathcal{N}=2$
vector multiplet \cite{Kapustin:2009kz,Jafferis:2010un}. The resulting
matrix model expression for the partition function will be given in Section \ref{sub:The-matrix-models}.
Specifically, the fluctuation determinant for an abelian vector multiplet
is $1$.

A supersymmetric defect which mimics the effects of computing the
supersymmetric R{\'e}nyi entropy can be introduced
by considering an additional background abelian vector multiplet
\[
\left\{ A_{\mu},\,\sigma^{A},\,D^{A},\,\text{fermions}\right\} .
\]
The defect configuration is of the form
\[
dA=\alpha\star\left[\gamma\right]_{D}\ ,\qquad\star D=i\alpha\left[\gamma\right]_{D}\wedge\left[\gamma\right]\ ,
\]
where $\left[\gamma\right]$ is the volume form on a maximal $\BS^1\subset \BS^{3}$,
and $\left[\gamma\right]_{D}$ is the Poincar{\'e} dual to this cycle.

For $\alpha=\frac{k}{n}$, this configuration can be imposed on a
dynamical vector multiplet by considering an additional vector multiplet
with connection $B$, an off-diagonal Chern-Simons coupling 
\[
S_{\text{BF}}^{\left(3d\right)}=\frac{in}{2\pi}\int_{\BS^{3}}\left(A\wedge dB+\sigma^{A}D^{B}+\sigma^{B}D^{A}+\text{fermions}\right),
\]
and a supersymmetric abelian Wilson loop of the form
\[
W^{\left(3d\right)}\left(k\right)\equiv\mbox{exp}\bigg[ik\oint_{\gamma}\left(B-i\left[\gamma\right]\sigma^{B}\right)\bigg].
\]
Localization reduces the above terms to\footnote{We universally denote the Lie algebra valued scalar zero mode as $\sigma$.}  
\begin{align}
\begin{aligned}
\exp\left(-S_\text{BF}^{\left(3d\right)}\right)\quad &\rightarrow\quad \exp\left(2\pi i\,n\,\sigma^{A}\sigma^{B}\right)\ ,\\
W^{\left(3d\right)}\left(k\right)\quad &\rightarrow\quad \exp\left(-2\pi k\,\sigma^{B}\right)\ ,
\end{aligned}
\end{align}
Integrating over the new modulus $\sigma^{B}$ then sets 
\begin{align}\label{3d_defect_shift}
\sigma^{A}=- i\,\frac{k}{n}\ .
\end{align}
Note that this vev for $\sigma^{A}$ is off the original contour of
integration and represents an imaginary mass term for chiral multiplets
to which the $A$ vector multiplet is coupled. We will argue later
that an imaginary Higgs type mass for a dynamical vector multiplet
can also be thought of this way. The imaginary mass is the entire
effect of the original defect on the localization computation. By
continuity, the same is true for an arbitrary $\alpha$. We will show
below that this continues to hold, for appropriate defects, in four
and in five dimensions.

\subsection{\label{sub:4d_defect}The $4d$ $\mathcal{N}=2$ codimension-two
defect}

We consider a $4d$ $\mathcal{N}=2$ theory consisting of vector multiplets
and hypermultiplets. We would like to show that the effect of inserting
a codimension-two defect is equivalent to the introduction of an
imaginary mass term.

\subsubsection{The $\mathcal{N}=2$ calculation on the four-sphere}

We begin by reviewing some aspects of the localization calculation
for an $\mathcal{N}=2$ theory on the round, and on the branched four-sphere
following \cite{Pestun:2007rz,Hama:2012bg,Huang:2014gca}. We refer
the reader to these papers for explicit actions and localizing terms.
The authors of \cite{Huang:2014gca} considered a smooth resolution
of the branched four-sphere, the resolved four-sphere, which is deformation
equivalent to it, and on which localization computations can be performed.
In this section, we retain an overall scale $\ell$, associated
with the size of the four-sphere, which will later be set to $1$.

We use the following coordinates on an $n$-fold covering of $\BS^{4}$
(the branched sphere)
\begin{align}
\begin{aligned}
ds^{2} & =\ell^{2}\left(d\theta^{2}+n^{2}\sin^{2}\theta\, d\tau^{2}+\cos^{2}\theta \,ds_{\BS_{2}}^{2}\right)\ ,\\
ds_{\BS_{2}}^{2} & =d\phi^{2}+\sin^{2}\phi \,d\chi^{2}\ .
\end{aligned}
\end{align}
We use the vielbein
\[
e^{1}=\ell\, d\theta\ ,\qquad e^{2}=n\ell\sin\theta\, d\tau\ ,\qquad e^{3}=\ell\cos\theta\, d\phi\ ,\qquad e^{4}=\ell\cos\theta\sin\phi\, d\chi\ ,
\]
and take a basis for the Clifford algebra
\[
\gamma_{a}=\begin{pmatrix}0 & \bar{\sigma}_{a}\\
\sigma_{a} & 0
\end{pmatrix},
\]
\[
\sigma_{i}\equiv-i\tau_{i}\ ,\qquad\bar{\sigma}_{i}=i\tau_{i}\ ,\qquad\sigma_{4}=\bar{\sigma}_{4}=\mathbbm{1}\ ,
\]
where $\tau_{i}$ are the Pauli matrices. 

An $\mathcal{N}=2$ supersymmetry is generated by a foursome of Weyl
spinors $\xi_{\alpha A},\bar{\xi}_{\dot{\alpha}A}$, subject to the
reality condition \cite{Hama:2012bg}
\[
\xi^{\alpha A}\equiv\left(\xi_{\alpha A}\right)^{\dagger}=\epsilon^{\alpha\beta}\epsilon^{AB}\,\xi_{\beta B}\ ,\qquad\bar{\xi}^{A\dot{\alpha}}\equiv\left(\bar{\xi}_{\dot{\alpha}A}\right)^{\dagger}=\epsilon^{\dot{\alpha}\dot{\beta}}\epsilon^{AB}\,\bar{\xi}_{\dot{\beta}B}\ .
\]
The subscript $A$ is an $SU(2)_{R}$ index. Indices $\alpha,\dot{\alpha}$
indicate a spinor transforming as a doublet under the left and right
$SU\left(2\right)$ factors of $\text{Spin}\left(4\right)$. Indices
$\alpha,\dot{\alpha},A$ are raised with $\epsilon^{\alpha\beta},\epsilon^{\dot{\alpha}\dot{\beta}},\epsilon^{AB}$
such that $\epsilon^{12}=1$ and $\epsilon_{AB}=-\epsilon^{AB}$. 

In order to preserve rigid supersymmetry, one must solve the Killing spinor equation. The relevant Killing spinor equation for the round/branched four-sphere
is 
\begin{align}
\begin{aligned}
\partial_{\mu}\xi_{A}+\frac{1}{8}{\omega_{\mu}}^{ab}\left(\sigma_{a}\bar{\sigma}_{b}-\sigma_{b}\bar{\sigma}_{a}\right)\xi-i{{A_{\mu}^{SU(2)_{R}}}_{A}}^{B}\xi_{B} &=-i\sigma_{\mu}\,\bar{\xi}'_{A}\ ,\\
\partial_{\mu}\bar{\xi}_{A}+\frac{1}{8}{\omega_{\mu}}^{ab}\left(\bar{\sigma}_{a}\sigma_{b}-\bar{\sigma}_{b}\sigma_{a}\right)\bar{\xi}_{A}-i{{A_{\mu}^{SU(2)_{R}}}_{A}}^{B}\bar{\xi}_{B} &=-i\bar{\sigma}_{\mu}\,\xi'_{A} \ ,
\end{aligned}
\end{align}
Here, we have introduced an $SU(2)_{R}$ background connection
$A^{SU(2)_{R}}$, and set all other supergravity background
fields besides the metric to zero. The solution is given by 
\[
A^{SU(2)_{R}}=\frac{n-1}{2}\begin{pmatrix}1 & 0\\
0 & -1
\end{pmatrix}d\tau\ ,
\]
and 
\begin{align}\label{eq:4d_Killing_spinors}
\begin{aligned}
\xi_{\alpha1}&=e^{-\frac{i}{2}\left(\tau+\chi\right)}\begin{pmatrix}-\sin\frac{\theta}{2}\cos\frac{\phi}{2}\\
\cos\frac{\theta}{2}\sin\frac{\phi}{2}
\end{pmatrix}\ ,&\qquad\xi_{\alpha2}&=e^{\frac{i}{2}\left(\tau+\chi\right)}\begin{pmatrix}\cos\frac{\theta}{2}\sin\frac{\phi}{2}\\
\sin\frac{\theta}{2}\cos\frac{\phi}{2}
\end{pmatrix}\ ,\\
{\bar{\xi}^{\dot{\alpha}}}_{1}&=e^{-\frac{i}{2}\left(\tau+\chi\right)}\begin{pmatrix}-\sin\frac{\theta}{2}\sin\frac{\phi}{2}\\
-\cos\frac{\theta}{2}\cos\frac{\phi}{2}
\end{pmatrix}\ ,&\qquad{\bar{\xi}^{\dot{\alpha}}}_{2}&=e^{\frac{i}{2}\left(\tau+\chi\right)}\begin{pmatrix}\cos\frac{\theta}{2}\cos\frac{\phi}{2}\\
-\sin\frac{\theta}{2}\sin\frac{\phi}{2}
\end{pmatrix}\ .
\end{aligned}
\end{align}
$\xi'$ can be extracted by contracting with $\sigma^{\mu}$/$\bar{\sigma}^{\mu}$.
In order to use the transformation generated by $\xi$ to perform
localization, one should impose an additional constraint to ensure
that the square of the transformation does not contain scale or $U(1)_{R}$
transformations 
\[
\xi^{\alpha A}\xi'_{\alpha A}=0\ ,\qquad\bar{\xi}^{\dot{\alpha}A}\bar{\xi'}_{\dot{\alpha}A}=0\ .
\]
This constraint has been taken into account above.

As shown in \cite{Hama:2012bg,Huang:2014gca}, one can introduce supergravity
backgrounds for the resolved four-sphere such that the same Killing
spinors are preserved. Moreover, one may use the same localizing term
for the vector and hypermultiplets in the presence of defects/squashing
as one does in the round sphere case. These localizing terms, which
are described in \cite{Pestun:2007rz,Hama:2012bg}, imply that all
components of a hypermultiplet must vanish on the moduli space. The
contour of integration for vector multiplet scalars compatible with the localizing terms is
\begin{equation}
\phi^{\dagger}=-\bar{\phi}\ ,\qquad\left(D^{AB}\right)^{\dagger}=-D_{AB}\ .\label{eq:4d_vector_contour}
\end{equation}
The localizing terms yield the following moduli space of zero modes
for a vector multiplet
\[
\phi=\bar{\phi}=-i\,\frac{\ell}{2}\,D_{12}=-i\,\frac{\sigma}{2}\ ,
\]
where $\sigma$ is a constant Lie algebra valued scalar. In addition,
there are point-like instantons localized at the north pole ($\theta=0,\phi=0$),
and anti-instantons at the south pole ($\theta=0,\phi=\pi/2$). The
complete matrix model expression for the partition function is given
in Section \ref{sub:The-matrix-models}.

\subsubsection{Introducing the defect}

We would like to introduce a surface defect into the computation on
the round sphere. The data for a surface defect can be embedded in
a background $\mathcal{N}=2$ abelian vector multiplet
\[
\left\{ A_{\mu},\,\phi,\,\bar{\phi},\,D_{AB},\,\lambda_A,\,\bar{\lambda}_A \right\}\ ,
\]
where $A_\mu$ is the gauge field, $\phi$ and $\bar{\phi}$ are complex scalar fields whose relationship depends on the contour of integration, $D_{AB}$ is an $SU\left(2\right)_R$ triplet of auxiliary scalars, and $\lambda_A,\bar{\lambda}_A$ independent Weyl fermions which are $SU\left(2\right)_R$ doublets.  
We want a supersymmetric defect, on the round four-sphere, supported
only on the two-sphere at $\theta=0$. Such a defect would have a field
strength
\[
F_{\text{defect}}=\alpha\,\delta\left(\theta\right)d\theta\wedge d\tau\ .
\]

In order to regularize the above configuration, we introduce a background
\[
A_{\text{defect}}=\alpha\, g_{\epsilon}\left(\theta\right)d\tau\ ,\qquad\alpha>0 \ ,
\]
such that\footnote{We write $g'_{\epsilon}\left(\theta\right)\equiv\partial_{\theta}g_{\epsilon}\left(\theta\right)$.}
\begin{equation}
F_{\text{defect}}=\alpha\, g'_{\epsilon}\left(\theta\right)d\theta\wedge d\tau\ .\label{eq:defect_field_strength}
\end{equation}
$g_{\epsilon}\left(\theta\right)$ is a smooth function satisfying
\[
\begin{cases}
g_{\epsilon}\left(\theta\right)=1\ , &\qquad \epsilon<\theta\le\frac{\pi}{2}\ ,\\
g_{\epsilon}\left(\theta\right)\sim\theta\ , & \qquad \theta\rightarrow0\ .
\end{cases}
\label{smoothing_function}
\]

The supersymmetry variation of the gaugino in an abelian vector multiplet
is \cite{Hama:2012bg} 
\begin{align}
\begin{aligned}
\delta\lambda_{A}&=\frac{1}{2}\sigma^{\mu\nu}F_{\mu\nu}\,\xi_{A}+2\slashed{D}\phi\,\xi_{A}+D_{AB}\,\xi^{B}+2\phi\,\slashed{D}\bar{\xi}_{A}\ ,\\
\delta\bar{\lambda}_{A}&=\frac{1}{2}\bar{\sigma}^{\mu\nu}F_{\mu\nu}\,\bar{\xi}_{A}+2\slashed{D}\bar{\phi}\,\bar{\xi}_{A}+D_{AB}\,\bar{\xi}^{B}+2\bar{\phi}\,\slashed{D}\xi_{A}\ .
\end{aligned}
\end{align}
We can complete \eqref{eq:defect_field_strength} to a supersymmetric
configuration by introducing additional bosonic backgrounds 
\[
D_{12}=-\alpha\,\frac{i}{\ell^{2}}\left(1-g_{\epsilon}\left(\theta\right)+\cot\theta g'_{\epsilon}\left(\theta\right)\right)\ ,\qquad\phi=\alpha\,\frac{g_{\epsilon}\left(\theta\right)-1}{2\ell} ,\qquad\bar{\phi}=\alpha\,\frac{g_{\epsilon}\left(\theta\right)-1}{2\ell}\ .
\]
In this background, the gaugino variations vanish with an arbitrary
spinor from \eqref{eq:4d_Killing_spinors}. In the limit $\epsilon\rightarrow0$
we get 
\[
g_{\epsilon}\left(\theta\right)\rightarrow1\ ,\qquad g'_{\epsilon}\left(\theta\right)\rightarrow\delta\left(\theta\right)\ ,
\]
\begin{equation}
A_{\text{defect}}=\alpha\, d\tau\ ,\qquad F_{\text{defect}}=\alpha\,\delta\left(\theta\right)d\theta\wedge d\tau\ ,\label{eq:vector_defect_configuration}
\end{equation}
\begin{equation}
D_{12}^{\text{defect}}=-\alpha\frac{i}{\ell^{2}}\frac{\delta\left(\theta\right)}{\theta}\ ,\qquad\phi=\bar{\phi}=0\ .\label{eq:auxilliary_D_defect_configuration}
\end{equation}
Note that both the smooth and singular configurations are off the
original contour of integration for the vector multiplet \eqref{eq:4d_vector_contour}
($D_{12}$ is imaginary).

\subsubsection{The supersymmetric Lagrange multiplier}

Following the logic of the $\mathbb{Z}_{n}$ gauge theory, we will
impose the configuration for the background vector multiplet using
a supersymmetric Lagrange multiplier. We make the vector multiplet
dynamical and introduce an $\mathcal{N}=2$ abelian tensor multiplet
$E$
\[
\left\{ E_{\mu\nu},\,G,\bar{G},\,L^{AB},\,\text{fermions}\right\} ,
\]
whose field strength sits in the linear multiplet $G$ \cite{deWit:2006gn}.\footnote{We use the letters $E$ and $G$ to indicate both the supermultiplets
and some of their components. We hope this does not cause too much
confusion.}
We couple $E$ to the vector multiplet (actually its field strength)
using a supersymmetric BF term \cite{deWit:2006gn}
\begin{equation}
S_{\text{BF}}^{\left(4d\right)}=\frac{in}{2\pi}\int\sqrt{g}\left(\phi\, G+\bar{\phi}\,\bar{G}-\frac{1}{2}D^{AB}L_{AB}+\frac{1}{4}\varepsilon^{\mu\nu\rho\sigma}E_{\mu\nu}F_{\rho\sigma}+\text{fermions}\right)\ .\label{eq:tensor_vector_action}
\end{equation}
We then introduce a Wilson surface operator for the tensor multiplet
by adding an additional copy of \eqref{eq:tensor_vector_action}, but
replacing all fields in the vector multiplet with their values from
\eqref{eq:vector_defect_configuration} and \eqref{eq:auxilliary_D_defect_configuration}, and with $\alpha=1$,
\begin{align}
W_{\text{surface}}\left(k\right) & \equiv\exp\left[\frac{ik}{2\pi}\int\sqrt{g}\left(-\frac{1}{2}D_{AB}^{\text{defect}}L^{AB}+\frac{1}{4}\varepsilon^{\mu\nu\rho\sigma}E_{\mu\nu}F_{\rho\sigma}^{\text{defect}}\right)\right] \ ,\label{eq:tensor_defect_action}\\
 & =\exp\left[ik\int_{\mathbb{S}_{\theta=0}^{2}}\sqrt{g}\left(i\ell^{2}L^{12}+\star_{2}E\right)\right]\ .
\end{align}
Integrating out the tensor multiplet would result in a supersymmetric
delta function setting all elements of the vector multiplet to their
values for the surface defect. Instead of doing this, we first perform
the localization.

\subsubsection{Localization of the tensor multiplet}

The localizing term for the abelian vector multiplet can be taken
from \cite{Hama:2012bg}. For the tensor multiplet, we must consider
the transformation of the fermions in this multiplet, denoted $\varphi^{A},\bar{\varphi}^{A}$.\footnote{We have decomposed each spinor $\varphi^{i}$ from \cite{deWit:2006gn}
as a doublet $\varphi^{A},\bar{\varphi}^{A}$. Note that the definition
of the Levi-Civita symbol in \cite{deWit:2006gn} contains an extra
factor of ``$i$'' compared to the usual definition: $\varepsilon^{1234}=1$.
Examination of the gravitino transformation in \cite{deWit:2006gn}
yields the following dictionary in relation to \cite{Hama:2012bg}
\begin{align}
\epsilon^{i}\rightarrow\begin{pmatrix}\xi^{A}\\
\bar{\xi}^{A}
\end{pmatrix}\ ,\quad\eta^{i}\rightarrow\frac{1}{2}\slashed{D}\begin{pmatrix}\xi^{A}\\
\bar{\xi}^{A}
\end{pmatrix}=-2i\begin{pmatrix}\xi'^{A}\\
\bar{\xi}'^{A}
\end{pmatrix}\ .
\end{align}
Superscripts on both sides indicate $SU(2)_{R}$ transformations.
Note the factor of $2$ difference with the $\eta$ used in e.g. \cite{Freedman:2012zz,Pestun:2014mja}.}
The variations are 
\begin{align}
\begin{aligned}
\delta\varphi^{A} & =\sigma^{\mu}\left(\partial_{\mu}L^{AB}+\frac{1}{2}\varepsilon^{AB}\varepsilon_{\mu\nu\rho\sigma}\partial^{\nu}E^{\rho\sigma}\right)\bar{\xi}_{B}-G\,\xi^{A}+2L^{AB}\,\eta_{B}\ ,\\
\delta\bar{\varphi}^{A} & =\bar{\sigma}^{\mu}\left(\partial_{\mu}L^{AB}+\frac{1}{2}\varepsilon^{AB}\varepsilon_{\mu\nu\rho\sigma}\partial^{\nu}E^{\rho\sigma}\right)\xi_{B}-\bar{G}\,\bar{\xi}^{A}+2L^{AB}\,\bar{\eta}_{B}\ .
\end{aligned}
\end{align}
We take the localizing term
\[
S_{\text{tensor localizing term}}=\int_{\mathbb{S}^{4}}\delta\left[\left(\delta\varphi\right)_{A}^{\dagger}\varphi^{A}+\left(\delta\bar{\varphi}\right)_{A}^{\dagger}\bar{\varphi}^{A}\right].
\]

We will use reality conditions for $G,\bar{G},L^{AB}$ appropriate
for the coupling to the vector multiplet fields \eqref{eq:tensor_vector_action}
\[
\bar{G}=-G^{\dagger}\ ,\qquad\left(L^{AB}\right)^{\dagger}=-L_{AB}\ ,
\]
which yields 
\[
\left(L^{11}\right)^{\dagger}=-L^{22}\ ,\quad\text{Im}\left(L^{12}\right)=0\ .
\]
We also define
\begin{align}
\begin{aligned}
H_{\mu}&\equiv\frac{1}{2}\varepsilon_{\mu\nu\rho\sigma}\partial^{\nu}E^{\rho\sigma}\ ,\\
G&=G_{r}+iG_{i}\ .
\end{aligned}
\end{align}
The bosonic part of the resulting localizing term is 
\begin{align}\label{eq:tensor_localizing_term} 
\begin{aligned}
S_{\text{tensor localizing term}} =\int_{\mathbb{S}^{4}}\Bigg[2&H^{\mu}H_{\mu}+2\left|\partial_{\mu}L^{12}\right|^{2}+2\left|\left(\partial_{\mu}+\frac{i}{\ell}v_{\mu}\right)L^{11}\right|^{2}\\
 & +\frac{1}{\ell^{2}}\left(8-2v^{\mu}v_{\mu}\right)\left|L^{11}\right|^{2}+2\left(G_{r}\right)^{2}+2\left(G_{i}+2\frac{L^{12}}{\ell}\right)^{2}\Bigg]\ .
\end{aligned}
\end{align}
The vector field $v_{\mu}$ is such that 
\[
\nabla^{\mu}v_{\mu}=0\ ,\qquad\left(8-2v^{\mu}v_{\mu}\right)>0\ .
\]
The localization locus is thus
\begin{equation}
H_{\mu}=0\ ,\qquad L^{11}=L^{22}=G_{r}=0\ ,\qquad G_{i}=-2\frac{L^{12}}{\ell}=G_{0}\ .\label{eq:tensor_localization_locus}
\end{equation}
with $G_{0}$ a real constant.

\subsubsection{Modification of the matrix model}

We will couple the vector multiplet representing the surface defect
to the physical vector multiplets and hypermultiplets of the $\mathcal{N}=2$
SCFT as detailed in Section \ref{sub:Coupling-to-the-n-copy-theory}. After
localization, the matrix model representing the $\BS^{4}$ partition
function is modified by the presence of the defect. We would like
to show that this modification amounts to giving the hypermultiplets
or the vector multiplets an imaginary mass, given by $-i\frac{k}{n}$.
Below we discuss all contributions of $V$ and $E$ to the matrix
model.

\paragraph{Classical contribution}

The multiplets $V$ and $E$ have a classical action given by \eqref{eq:tensor_vector_action}
and \eqref{eq:tensor_defect_action}. After localization, these terms
give insertions in the matrix model. Using the vevs for the scalar
fields
\[
\phi=\bar{\phi}=-i\,\frac{\sigma}{2}\ ,\qquad D_{12}=\frac{\sigma}{\ell} \ ,
\]
and 
\[
G=\bar{G}=i\,G_{0}\ ,\qquad L^{12}=-\frac{\ell}{2} G_{0}\ ,
\]
we get
\begin{align}
\begin{aligned}
\exp\left(-S_{\text{BF}}^{\left(4d\right)}\right)\quad &\rightarrow\quad \exp\left(-2\pi i \ell^{4}\,n\,\sigma\, G_{0}\right)\ ,\\
W_{\text{surface}}\left(k\right)\quad &\rightarrow\quad \exp\left(2\pi\ell^{3}\,k\,G_{0}\right)\ .
\end{aligned}
\end{align}

\paragraph{Perturbative contributions}

The quadratic approximation to the localizing term \eqref{eq:tensor_localizing_term} for the tensor multiplet around the locus
\eqref{eq:tensor_localization_locus} is independent of $G_{0}$. The
one-loop determinant is thus a $G_{0}$ independent number, which
is furthermore equal to the partition function of a free tensor multiplet
on the round four-sphere. 
This multiplet can be dualized into an uncharged
massless hypermultiplet whose partition function can be deduced from
the expressions in \eqref{sub:Perturbative_contributions}. The one-loop determinant for $A$ is trivial.

\paragraph{Non-perturbative contributions}

In the absence of two-cycles, the equation 
\[
H=\star dE=0 \ ,
\]
implies that we can set 
\[
E=0\ ,
\]
up to tensor gauge transformations. When there are non-trivial two-cycles,
the form $E$ is closed and the tensor gauge transformations imply
that $E$ is gauge equivalent to $0$ whenever it represents an integral
class. The remaining moduli of $E$ can be identified with the values
of the possible Wilson surfaces. The gauge field $A$ also has moduli
in this situation. A smooth instanton configuration for $A$ may exist,
where $A$ is a non-trivial connection with integral flux on the two-cycle.

The analysis above applies only to smooth configurations for $A,E$
and their gauge transformation parameters. Experience shows that we
should allow singular configurations for $A$, as we did for the non-abelian
gauge fields. The coupling of $E$ to the field strength $F_{A}$
means that we should consider allowing singular configurations for
$E$ as well, at least at the poles. In fact, the configuration \eqref{eq:vector_defect_configuration}
which we are trying to reproduce has a singularity on the entire maximal
two-sphere. If we allow $E$ to have the singularities at the same
position as $A$, then the coupling 
\[
\frac{in}{2\pi}\int F_{A}\wedge E \quad\subset\quad \frac{in}{2\pi}\int V\,G\ ,
\]
induces a non-vanishing classical contribution upon localization
\[
\frac{in}{2\pi}\int F_{A}\wedge E\quad\rightarrow\quad\frac{in}{2\pi}\int F_{A}^{\text{inst}}\wedge E^{\text{classical}}\ .
\]
The form $E$ is again closed and the tensor gauge transformations
again imply that $E$ is gauge equivalent to $0$ whenever it represents
an integral class. The remaining moduli of $E$, parameterizing singular
closed two-forms localized at the poles modulo integral forms, are
angular parameters. Upon localization, there is a contribution to
the matrix model of the form 
\[
n\int F_{A}\wedge E \quad\rightarrow\quad n\int F_{A}^{\text{inst}}\wedge\left[\alpha\right]\ ,
\]
where $F_{A}^{\text{inst}}$ is the field strength of some singular
instanton configuration at the poles and $\left[\alpha\right]$ is
a representative of any singular two-form at the poles, which is
defined only up to integral classes. The part of the electric defect
containing an integral over $E$ is, however, trivial
\[
\oint_{\partial\Sigma}E=\oint_{\Sigma}H=0\ ,
\]
because $E$ is being paired with a trivial cycle represented by $\partial\Sigma$.
Integration of 
\[
\exp\left(in\int F_{A}^{\text{inst}}\wedge\left[\alpha\right]\right)\ ,
\]
over all $\alpha$ then restricts $F_{A}^{\text{inst}}$ to vanish.
This means that we only need to work in the instanton number $0$
sector for $A$.

\paragraph{Integrating out }

In addition to the above, there is a remaining integration over $G_{0}$.
This integral, and the one over $\sigma$, can be done explicitly.
The result is simply to set 
\[
\sigma~\rightarrow~ -\frac{i}{\ell}\frac{k}{n}\ ,\label{4dshift}
\]
or in the units of the rest of the sections
\[
\sigma=-i\,\frac{k}{n}\ .
\]
This was what we set out to show. 

The parameter entering the one-loop and instanton contributions in
the matrix model, described in Section \ref{sec:Squashing_from_defects}, is $im$,
where $m$ is identified with the $\sigma$ of a background vector
multiplet. Hence, our surface operator is equivalent to an imaginary
mass 
\[
m=-i\,\frac{k}{n}\ ,
\]
equivalently, a shift of the parameters by 
\[
q_{\text{surface}}=\frac{k}{n}\ . \label{4dsurfacecharge}
\]

\subsection{The 5$d$ $\CN=1$ codimension-two defect}\label{sub:5d_defect} 

We describe the implementation
of codimension-two defect operators in five-dimensional $\CN=1$
theories. Most of the analysis is similar to the four-dimensional
case, so we will be brief. Conventions and notations are the same as
in \cite{Hama:2014iea}.

\subsubsection{Killing spinors on the five-sphere}

Supersymmetric field theories on curved spaces are systematically
obtained in the rigid limit of the $\CN=1$ supergravity in five dimensions
which has an $SU(2)_{R}$ symmetry whose indices are denoted by $A,B$ as in the 4$d$ case.
There are $SU(2)_{R}$ gauge field
$V_{\mu}^{AB}$, $SU(2)_{R}$ triplet scalar field $t^{AB}$ and the
other fields which are irrelevant to the following discussion in the
Weyl multiplet \cite{Kugo:2000hn}. The variations of the fermions
in the multiplet have to vanish for preserving supersymmetries on
given background fields and the solutions are the Killing spinors.
We set the radius of the five-sphere to one from the beginning to simplify the discussion.

The round unit five-sphere allows the Killing spinor in the coordinates
\begin{align}
ds_{\BS^{5}}^{2}=d\theta^{2}+\sin^{2}\theta\,d\tau^{2}+\cos^{2}\theta\,ds_{\BS^{3}}^{2}\ ,\label{RoundS5}
\end{align}
when the background fields are set to \cite{Hosomichi:2012ek,Hama:2014iea}
\begin{align}
t_{~B}^{A}=\frac{1}{2}(\sigma_{3})_{~B}^{A}\ ,\qquad V_{~B}^{A}=-\frac{i}{2}(\sigma_{3})_{~B}^{A}\,d\tau\ ,\qquad\text{others}=0\ .
\end{align}
With the hermitian gamma matrices in tensor product forms 
\begin{align}
\Gamma^{1}=\sigma^{1}\otimes{\bf 1}_{2}\ ,\qquad\Gamma^{2}=\sigma^{2}\otimes{\bf 1}_{2}\ ,\qquad\Gamma^{i+2}=\sigma^{3}\otimes\sigma^{i}\ ,\quad(i=1,2,3)\ ,
\end{align}
a spinor $\xi^{A}$ in five dimensions is also written as tensor products
of spinors $\xi^{A}$ and $\eta^{A}$ in two and three dimensions
\begin{align}
\xi^{A}=\xi^{A}\otimes\eta^{A}\ .
\end{align}
The Killing spinor on the round sphere (\ref{RoundS5}) is given by
\begin{align}
\begin{aligned}\xi^{1} & =\left(e^{\frac{i}{2}\theta\,\sigma_{1}}\zeta^{1}\right)\otimes\eta_{+}\ ,\qquad & \sigma_{3}\zeta^{1} & =\zeta^{1}\ ,\\
\xi^{2} & =\left(e^{-\frac{i}{2}\theta\,\sigma_{1}}\zeta^{2}\right)\otimes\eta_{-}\ ,\qquad & \sigma_{3}\zeta^{2} & =-\zeta^{2}\ ,
\end{aligned}
\end{align}
where $\zeta^{1,2}$ are constant spinors and $\eta_{\pm}$ are the
Killing spinors on a unit three-sphere 
\begin{align}
\left(\partial_{i}+\frac{i}{2}\sigma_{i}\right)\eta_{\pm}=\pm\frac{i}{2}\sigma_{i}\,\eta_{\pm}\ .
\end{align}

\subsubsection{Codimension-two defects}

A surface defect in a flavor symmetry is specified as a singular configuration
in a background $\CN=1$ abelian vector multiplet 
\begin{align}
\{A_{\mu},\,\sigma,\,Y^{AB},\,\lambda^{A}\}\ ,
\end{align}
where  $A_{\mu}$ the gauge
field, $\sigma$ a real scalar, $Y^{AB}$ an $SU(2)_{R}$ triplet
scalar and $\lambda^{A}$ an $SU(2)_{R}$-Majorana fermion, respectively.

We want a codimension-two surface defect at $\theta=0$ on the round
sphere. Introducing a smoothing function $g_{\epsilon}(\theta)$ as
in the four-dimensional case (\ref{smoothing_function}), the background
gauge field describing the defect is 
\begin{align}
A_{\text{defect}}=\alpha\,g_{\epsilon}(\theta)d\tau\ ,
\end{align}
whose field strength becomes 
\begin{align}
F_{\text{defect}}=\alpha\,g_{\epsilon}'(\theta)\,d\theta\wedge d\tau\ .
\end{align}

This configuration is supersymmetric if the real scalar $\sigma_{\text{defect}}$,
the gaugino $\lambda_{\text{defect}}^{A}$ and the triplet scalar
$Y_{\text{defect}}^{AB}$ take the following forms 
\begin{align}
(Y_{\text{defect}})_{~B}^{A}=-\frac{i}{2}\alpha\,g'_{\epsilon}(\theta)(\sigma_{3})_{~B}^{A}\ ,\qquad\sigma_{\text{defect}}=\lambda_{\text{defect}}^{A}=0\ .
\end{align}
This configuration is invariant under the supersymmetric transformation,
especially one sees 
\begin{align}
0=\delta\lambda_{\text{defect}}^{A}=\frac{1}{4}\Gamma^{\mu\nu}\xi^{A}(F_{\text{defect}})_{\mu\nu}+(Y_{\text{defect}})_{~B}^{A}\,\xi^{B}\ .
\end{align}
One can replace $g'_{\epsilon}(\theta)$ with the delta function $\delta(\theta)$
in the singular limit $\epsilon\to0$.

\subsubsection{Defects and supersymmetric Lagrange multiplier}

We will implement defects in a flavor symmetry by coupling the abelian
vector multiplet to the linear multiplet 
\begin{align}
\{L^{AB},\,E_{\mu\nu},\,N,\,\,\varphi^{A}\}\ ,
\end{align}
consisting of an $SU(2)_{R}$ triplet scalar $L^{AB}$, an antisymmetric
tensor gauge field $E_{\mu\nu}$, a real scalar $N$ and an $SU(2)_{R}$-Majorana
fermion $\varphi^{A}$, through a supersymmetric BF coupling \cite{Kugo:2000hn}
\begin{align}
S_{\text{BF}}^{(5d)}=\frac{in}{\pi}\int\sqrt{g}\left(\frac{1}{4}F_{\mu\nu}E^{\mu\nu}+Y_{AB}L^{AB}-\frac{1}{2}\sigma N+2i\lambda\varphi\right)\ .\label{BF5d}
\end{align}
It is gauge invariant for an integer $n$ as $E_{\mu\nu}$ transforms
as $\delta_{\text{gauge}}E^{\mu\nu}=\partial_{\rho}\Lambda^{\mu\nu\rho}$.
The supersymmetric transformation laws of the linear multiplet with
only the backgrounds $t^{AB}$ and $V^{AB}$ turned on are given by
\begin{align}
\begin{aligned}\delta L^{AB} & =2i\xi^{(A}\varphi^{B)}\ ,\\
\delta E^{\mu\nu} & =2i\xi\Gamma^{\mu\nu}\varphi\ ,\\
\delta N & =-2\xi\Gamma^{\mu}(D_{\mu}\varphi)-10i\xi^{A}\varphi^{B}t_{AB}
\ ,\\
\delta\varphi^{A} & =i(D_{\mu}L^{AB})\Gamma^{\mu}\xi_{B}-4t_{~C}^{B}L_{~B}^{C}\xi^{A}-6t_{~C}^{(A}L^{B)C}\xi_{B}+\frac{i}{2}\Gamma_{\mu}\xi^{A}\partial_{\nu}E^{\mu\nu}+\frac{1}{2}\xi^{A}N
\ .
\end{aligned}
\label{LM_susy-1}
\end{align}

In addition, we introduce a surface operator for the linear multiplet
with the BF coupling to the defect abelian vector multiplet
\begin{align}
\begin{aligned}W_{\text{surface}}(k) & =\exp\left[-\frac{ik}{2\pi}\int\sqrt{g}\,\left(-(Y^{\text{defect}})_{~A}^{B}\,L_{~B}^{A}+\frac{1}{4}F_{\mu\nu}^{\text{defect}}E^{\mu\nu}\right)\right]\ ,\\
 & =\exp\left[-ik\,\alpha\int_{\BS_{\theta=0}^{3}}\left(-\frac{i}{2}\,(L_{~1}^{1}-L_{~2}^{2})\,\omega_{3}+\frac{1}{2}\ast E\right)\right]\ ,
\end{aligned}
\label{Surface5d}
\end{align}
where $\omega_{3}$ is the volume form of the unit three-sphere and
$\ast$ is the Hodge operator in five dimensions. 
Choosing $\alpha=2$ the defect operator is gauge invariant for an integer $k$.

The linear multiplet plays a role of the Lagrange multiplier and integrating
it out in the path integral with the BF term and the Wilson surface
results in setting the vector multiplet to the defect configuration.

Instead of doing so, we localize the tensor multiplet on the round
sphere with the localizing term 
\begin{align}
V_{\text{tensor}}^{\text{(loc)}}=\delta\left[\left(-i(D_{\mu}L_{~B}^{A})\xi^{B}\Gamma^{\mu}-4t_{~C}^{B}L_{~B}^{C}\xi^{A}-6t_{~C}^{(A}L^{B)C}\xi_{B}-\frac{i}{2}(\partial_{\nu}E^{\mu\nu})\xi^{A}\Gamma_{\mu}+\frac{1}{2}N\xi^{A}\right)\varphi_{A}\right]\ ,
\end{align}
whose bosonic part is written as
\begin{align}
\begin{aligned}\frac{1}{2}(D_{\mu}L_{~B}^{A})(D^{\mu}L_{~A}^{B}) & +\frac{1}{4}(\partial^{\nu}E_{\mu\nu})(\partial_{\nu}E^{\mu\nu})+\frac{1}{4}(N-8\hat{L}_{~A}^{A})^{2}+18\,\hat{L}^{(AB)}\hat{L}_{(AB)}\ ,
\end{aligned}
\end{align}
where we defined $\hat{L}^{AB}\equiv t_{~C}^{A}L^{CB}$.

In localizing the tensor multiplet, we impose the reality condition
for $L_{~B}^{A}$ 
\begin{align}
(L_{~B}^{A})^{\dagger}=L_{~A}^{B}\ ,
\end{align}
that is equivalent to 
\begin{align}
(\hat{L}_{~A}^{A})^{\dagger}=\hat{L}_{~A}^{A}\ ,\qquad(\hat{L}^{(AB)})^{\dagger}=L_{(AB)}\ .
\end{align}
This choice makes the bosonic part of the localizing term be semi-positive
definite and the tensor multiplet localizes to 
\begin{align}
D_{\mu}L_{~B}^{A} & =0\ ,\qquad\partial_{\nu}E^{\mu\nu}=0\ ,\qquad N=8\hat{L}_{~A}^{A}\ ,\qquad\hat{L}_{(AB)}=0\ ,
\end{align}
which on the round sphere yields 
\begin{align}
N=4(L_{~1}^{1}-L_{~2}^{2})\equiv N_{0}=\text{const}\ ,\qquad L_{~2}^{1}=L_{~1}^{2}=0\ .
\end{align}

Localizing the vector multiplet to the fixed locus 
\begin{align}
A_{\mu}=0\ ,\qquad Y_{AB}=0\ ,\qquad\sigma =\text{const}\ ,
\end{align}
the BF term (\ref{BF5d}) and the surface operator (\ref{Surface5d})
end up with 
\begin{align}
\begin{aligned}\exp\left(-S_{\text{BF}}^{(5d)}\right)\quad & \rightarrow\quad\exp\left(\frac{i\pi^{2}}{2}n\sigma N_{0}\right)\ ,\\
W_{\text{surface}}(k)\quad & \rightarrow\quad\exp\left(-\frac{\pi^{2}}{2}kN_{0}\right)\ .
\end{aligned}
\end{align}
the integration over $N_{0}$ sets 
\begin{align}
\sigma =- i\,\frac{k}{n}\ .\label{5d_defect_shift}
\end{align}
This value is consistent with the mass shift by the parameter \eqref{4dsurfacecharge} as in the 4$d$ case as we will confirm in Section \ref{sec:Squashing_from_defects}.

\section{Squashing from defects}
\label{sec:Squashing_from_defects}

We are now in a position to demonstrate the relationship between the
supersymmetric R{\'e}nyi entropy and supersymmetric
codimension-two defects. In the original setup, the supersymmetric
R{\'e}nyi entropy is defined as 
\begin{align}
S_{n}^{\text{susy}}\equiv\frac{1}{1-n}\log\left|\frac{Z_{n}^{\text{susy}}}{\left(Z_{1}\right){}^{n}}\right|\ ,\label{SRE_def-1}
\end{align}
where $Z_{n}$ is the appropriate supersymmetry preserving partition
function on the branched $d$-sphere, \emph{or} the squashed $d$-sphere.
The schematic form of the localization calculation for $Z_{n}$ is
\[
Z_{n}=\sum\int_{\text{moduli}}\left[Z_{n}^{\text{classical}}\left(\text{moduli}\right)Z_{n}^{\text{pert}}\left(\text{moduli}\right)\right],
\]
where the sum/integral is over the moduli space of supersymmetric
zero modes, $Z_{n}^{\text{classical}}$ is the exponential of minus
the Euclidean action evaluated on the moduli space, and $Z_{n}^{\text{pert}}$
is the one-loop determinant obtained by evaluating the path integral
over non-zero modes in the quadratic approximation around the moduli
space.

The partition function for the $n$-copy theory on the round sphere
is simply 
\[
\left(Z_{1}\right)^{n}.
\]
We would like to demonstrate that $Z_{n}$ can be computed using $Z_{1}$
and the insertion of defects, i.e. 
\begin{equation}
Z_{n}=Z_{n}^{\text{defect}}.\label{eq:defect_calculation}
\end{equation}
Where the right hand side is the partition function of the $n$-copy theory on
the round sphere in the presence of a specific codimension-two defect
described in the next subsection. After the linear field redefinition,
the result is 
\[
Z_{n}^{\text{defect}}\equiv\sum\int_{n\text{ moduli}}\left[\delta_{\text{moduli}}\prod_{k=0}^{n-1}Z_{1}^{\text{classical}}\left(\text{moduli}\right)Z_{1}^{\text{pert-defect}}\left(k,\text{moduli}\right)\right].
\]
The outer sum/integral is over $n$ copies of the moduli. The symbol
$\delta_{\text{moduli}}$ is a placeholder for the sewing operation
which identifies how the squashed sphere moduli space fractionalizes
into $n$ copies. $Z_{1}^{\text{classical}}$ is the classical contribution
of a single copy of the theory evaluated on the round sphere. $Z_{1}^{\text{pert-defect}}\left(k,\text{moduli}\right)$
is the perturbative contribution, in the quadratic approximation around
the moduli space, in the presence of the appropriate defect for the
$k^{th}$ copy.

\subsection{\label{sub:Coupling-to-the-n-copy-theory}Coupling defects to the
$n$-copy theory}

The superfield $V$, which carries the information about the supersymmetric defect,
 must be coupled to the physical fields of the
$n$-copy theory. For matter multiplets, this is the usual minimal
coupling of $V$ to chiral multiplets or hypermultiplets. We need
not consider the effect of the defect on the other terms in the action
involving matter fields, since these vanish at the level of the quadratic
approximation around the localization locus for any of the setups
we consider. The coupling of a flat connection carried by $V$ to
physical gauge fields can be accomplished by formally performing the
field redefinition in the introduction. Since non-abelian gauge fields
are not free fields at finite gauge coupling, the resulting action
would inevitably look like a gauge non-invariant mess. However, gauge
invariance, with the caveats already mentioned, is guaranteed by the
ability to undo the field redefinition. 

If the gauge group is $SU(N)$, there is a physical procedure
which implements the right defect and makes clear the form of the
coupling to a background vector, at the level of the quadratic approximation
to the moduli space. We use the language of $4d$ $\mathcal{N}=2$,
but the same applies to any of the theories under consideration. 

First, consider the enlarged gauge group $U(nN)$, where
the gauge group of the $n$-copy theory, $SU(N)^{n}$,
is embedded as a block diagonal subgroup
\[
\left(\begin{array}{cccc}
SU(N)_{0} & 0 & ~\cdots & 0\\
0 & SU(N)_{1} & ~\cdots & 0\\
\vdots & \vdots & \ddots & ~\vdots\\
0 & 0 & ~\cdots & SU(N)_{n-1}
\end{array}\right).
\]
The required codimension-two defect, before the field redefinition,
can be viewed as a flat connection, on $\BS^d\backslash\partial\Sigma$,
which is represented by a one-form with holonomy
\begin{equation}
\left(\begin{array}{ccccc}
0 & ~1~ & ~0~ & \cdots &0\\
0 & 0 & 1 & \cdots & 0\\
\vdots & \vdots && \ddots & \vdots\\
0 & 0 & 0 & \cdots & 1\\
1 & 0 & 0 & \cdots & 0
\end{array}\right)\ ,\label{eq:defect_holonomy}
\end{equation}
along any path encircling $\partial\Sigma$ once in a chosen direction.
To preserve supersymmetry, we consider a Gukov-Witten type surface
defect with the holonomy \eqref{eq:defect_holonomy} specifying the
data \cite{Gukov:2006jk,Gukov:2014gja}. 

To go back to the $n$-copy theory, while keeping the defect, one
should first Higgs $U(nN)$ down to $SU(N)^{n}$
by giving an appropriate large vev to the adjoint scalar in the vector
multiplet
\[
X^{\text{Higgs}}=
\left(\begin{array}{cccc}
t_{0} & 0 & ~\cdots & 0\\
0 & t_{1} & ~\cdots & 0\\
\vdots & \vdots & ~\ddots & \vdots\\
0 & 0 & ~\cdots & t_{n-1}
\end{array}\right).
\]
Upon taking $t_{i}\rightarrow\infty$, all modes not coming from the
original $n$-copy theory are infinitely massive and do not contribute
to the computation. $X^{\text{Higgs}}$ should really be considered
only up to permutations of the $t_{i}$, which are a part of the Weyl
group of the theory. In fact, the holonomy \eqref{eq:defect_holonomy}
acting on $X^{\text{Higgs}}$ produces such a permutation. Although
we do not show this explicitly, we take this to mean that the Gukov-Witten
operator with this data preserves the same supersymmetry as $X^{\text{Higgs}}$.
After a change of variables, which is in this case a constant $SU(nN)$
gauge transformation, the fields in the vector multiplet for $SU(N)_{k}$
acquire a monodromy $\exp\left(2\pi ik/n\right)$ around $\partial\Sigma$.\footnote{Note that the diagonal elements of $SU(N)_{k}$ also acquire
this monodromy. Had we tried to implement the defect using a Gukov-Witten
type surface defect in each $SU(N)_{k}$, this would not
have been so.}

The physical effect of the defect on vector multiplets can now be
examined more carefully at the level of the quadratic approximation
to the moduli space of the remaining light fields. Since the action
in this approximation is quadratic, the monodromy can be traded for
a coupling to a background vector multiplet with a specific profile.
This profile is singular, and determined by the value of the monodromy
and by supersymmetry. It is this multiplet, denoted by $V$, which
arises in our realization of the supersymmetric $\mathbb{Z}_{n}$
gauge theory.

\subsection{Mass terms and vortices}

A vev for the scalar components of $V$ appears as a supersymmetric
mass term. For vector multiplets, this is a mass term of the Higgs
type. For matter multiplets, it is a mass term associated with the
$U(1)$ flavor symmetry of a free hypermultiplet or chiral
multiplet. As shown in Section \ref{sec:Defects_from_abelian_duality}, the vortex charge is equivalent, in the matrix model, to an imaginary
mass term.
Any required vortex charge can be produced using the supersymmetric
codimension-two defects of Section \ref{sec:Defects_from_abelian_duality}.\footnote{We have shown that this is so for any fraction $\frac{k}{n}$. By
continuity, the same is true for any real number.} The supersymmetric $\mathbb{Z}_{n}$ gauge theory naturally produces
vortex charges of the type $k/n$. One still needs to show that taking
these vortex charges reproduces the partition function on the squashed
sphere. We show this for each element of the matrix model in the following
subsections. 

To reproduce the squashed sphere result using defects, one must correctly
choose the origin of the imaginary part of the mass deformation parameters
in both the perturbative and non-perturbative parts. As shown in \cite{Okuda:2010ke},
there is a subtlety in doing so for the hypermultiplet mass of the
$4d$ $\mathcal{N}=2^{*}$ theory. Only by choosing this origin correctly
can one produce a Gaussian matrix model for Wilson loops in $\mathcal{N}=4$,
which was the topic of \cite{Pestun:2007rz}. The correct origin for
the mass deformation is associated with enhanced supersymmetry, and
with the vanishing of the instanton contributions for $\mathcal{N}=4$.
As shown in \cite{Okuda:2010ke}, for the $\mathcal{N}=2^{*}$ theory,
the mass parameter usually used in the Nekrasov partition function
should be shifted as\footnote{In \cite{Okuda:2010ke}, a mass deformation parameter $m$ appears.
It should be identified with $m_{f}$ in our notation. } 
\[
m_{f}~\rightarrow~ m_{f}+\frac{1}{2}\left(\epsilon_{1}+\epsilon_{2}\right)\ ,
\]
in order to align it with the mass parameter used in the perturbative
part. This shift comes from careful examination of the equivariant
action of the square of the supercharge used in the localization,
and so it should apply to any $\mathcal{N}=2$ theory. In terms of
the squashing parameters $\omega_{1,2}$ it is 
\[
m_{f}~\rightarrow~ m_{f}+\frac{1}{2}\left(\omega_{1}+\omega_{2}\right)\ ,
\]
which applies equally well at the north and south poles. The analogous
statement in the 5$d$ $\mathcal{N}=1^{*}$ theory is
\[
m_{f}~\rightarrow~ m_{f}+\frac{1}{2}\left(\omega_{1}+\omega_{2}+\omega_{3}\right)\ .
\]

In addition to aligning the mass parameters, one must determine the
value of $m_{f}$ at the conformal fixed point. According to \cite{Okuda:2010ke},
the round four-sphere ($\epsilon_{1}=\epsilon_{2}$) value, after
the shift, is 
\[
m_{f}=0\ .
\]
At this value, one recovers a Gaussian matrix model for $\mathcal{N}=2^{*}$,
which is interpreted as a signal of supersymmetry enhancement to $\mathcal{N}=4$.
On the squashed sphere, one must take \cite{Hosomichi:2016flq} 
\[
m_{f}=\pm\frac{1}{2}\left(\epsilon_{1}-\epsilon_{2}\right)\ ,
\]
to achieve the same effect . In terms of the squashing parameters,
this is 
\[
m_{f}=\pm\frac{1}{2}\left(\omega_{1}-\omega_{2}\right)\ .\label{4d_mass_shift}
\]
We will choose the upper sign.

For the 5$d$ $\mathcal{N}=1^{*}$ model, the prescription described in
\cite{Kim:2011mv,Bullimore:2014upa} is to shift the mass term by an additional amount
\[
\pm\frac{1}{2}\left(\omega_{1}+\omega_{2}-\omega_{3}\right)\ .\label{5d_mass_shift}
\]
This choice was made so as to make contact with the 4$d$ results in
\cite{Okuda:2010ke}. We will, instead, use an additional shift by 
\[
-\frac{1}{2}\left(\omega_{2}+\omega_{3}-\omega_{1}\right)\ ,\label{5d_mass_shift_new}
\]
which has the property that it reproduces the 4$d$ result for the instanton partition function after taking $\beta\rightarrow 0$.

Combining the shifts gives the following mass parameter for the instanton
partition function
\begin{itemize}
\item For the 4$d$ calculation $m_{f}=\omega_{1}$, which corresponds to $m_{f}=\epsilon_{1},\epsilon_{2}$
at the north and south pole respectively. 
\item For the 5$d$ calculation $m_{f}=\omega_{1}$, which corresponds to $m_{f}=\epsilon_{1},\epsilon_{2},\frac{2\pi}{\beta}$
at the three fixed points. 
\end{itemize}
We will denote these values, in any dimension, as $m_{f}^{0}$. Note
that the one-loop part does not require the first shift, so its effective
mass parameter is different. 

A codimension-two defect operator, with charge $\frac{k}{n}$, further
shifts the mass parameter by 
\[
m_{f}\rightarrow m_{f}+\frac{k}{n}\ .
\]
We will show that such shifts, which can be interpreted in terms of
the $\mathbb{Z}_{n}$ gauge theory, are enough to reproduce the supersymmetric
R{\'e}nyi entropy in four and five dimensions. Had
we not implemented the second shift above, a different vortex charge
would have been required for the hypermultiplets. This additional
vortex charge can be thought of as an additional twist arising from
the R-charge of a hypermultiplet. It can be incorporated by adjusting
the holonomy \eqref{eq:defect_holonomy} to include an overall phase associated with
$U(1)\subset U(nN)$. The result in three dimensions
does not include instantons and the vortex charge is 
\[
q_{k}^{\text{3$d$ vortex}}=\frac{\Delta}{2}\left(\frac{1}{n}-1\right)+\frac{k}{n}\ ,
\]
where $\Delta$ is the R-charge of the chiral multiplet. A vector
multiplet can be interpreted as a chiral multiplet with $\Delta=0$.
The addition of a term proportional to $\Delta$ was interpreted in
\cite{Nishioka:2013haa} as an additional R-symmetry twist of the
usual R{\'e}nyi entropy, which is needed to make
it supersymmetric. It should be noted that, if all chiral multipets
have the same $\Delta$, this term too can be canceled by shifting
the origin of the mass term for chiral multiplets in the imaginary
direction, as explained above. The result is a supersymmetric observable,
with or without this additional shift. We do not currently know of
a criterion which makes one of these options more relevant.

\subsection{\label{sub:The-matrix-models}The matrix models}

We collect the expressions for the matrix models associated to the
squashed sphere partition functions in three, four and five dimensions.
Our conventions for integration over the Lie algebra are in Appendix \ref{sec:Conventions}.
Special functions in the one-loop determinants and instanton contributions
are defined in Appendices \ref{sec:Special-functions} and \ref{sec:The-instanton-partition},
respectively. 

The matrix model for a 3$d$ $\mathcal{N}=2$ theory on the squashed
sphere is \cite{Hama:2011ea,Imamura:2011wg} 
\begin{align}\label{3dMM}
\begin{aligned}Z^{\text{susy 3$d$}}=\frac{1}{|W|}\int\prod_{i=1}^{\text{rank}\,G}\frac{d\sigma_{i}}{\sqrt{\omega_{1}\omega_{2}}}\,e^{\frac{\pi i\kappa}{\omega_{1}\omega_{2}}\,\text{Tr}(\sigma^{2})}\cdot & \prod_{\alpha\in\D_{+}}S_{2}\left(i\alpha(\sigma)|\boldsymbol{\omega}\right)S_{2}\left(-i\alpha(\sigma)|\boldsymbol{\omega}\right)\\
 & \cdot\prod_{I=1}^{\text{\# of chirals}}\prod_{\rho\in\CR_{I}}S_{2}\left(i\rho(\sigma)+im_{I}+\frac{|\boldsymbol{\omega}|}{2}\Delta_{I}\bigg|\boldsymbol{\omega}\right)^{-1}\ ,
\end{aligned}
\end{align}

\begin{itemize}
\item $\sigma_{i}$ runs over the Cartan of the Lie algebra $\mathfrak{g}$
of the group $G$, which we have assumed is $U(N)$. Our
conventions are such that $\sigma_{i}$ are real (see Appendix \ref{sec:Conventions}).
\item $\Delta_{+}$ are the positive roots of $\mathfrak{g}$. $\rho$ denotes
a weight in the representation $\mathcal{R}_{I}$ associated to the
$I^{th}$ chiral multiplet.
\item $\left|W\right|$ is the size of the Weyl group.
\item $\Delta_{I}$ is the R-charge of the $I^{th}$ chiral multiplet.
\item $\kappa$ is the Chern-Simons level. 
\item $\boldsymbol{\omega}=\left(\omega_{1},\omega_{2}\right)$ are squashing
parameters, and 
\[
\left|\boldsymbol{\omega}\right|\equiv\left|\sum\omega_{i}\right|=\left|\omega_{1}+\omega_{2}\right|.
\]
The supersymmetric R{\'e}nyi entropy is computed
using $\boldsymbol{\omega}=\left(1,\frac{1}{n}\right)$.
\item We will set the mass parameters $m_{I}$ to zero.
\end{itemize}
The matrix model for a 4$d$ $\mathcal{N}=2$ theory on the squashed
sphere is \cite{Hama:2012bg}

\begin{align}\label{4dMM}
\begin{aligned}Z^{\text{susy 4$d$}}=\frac{1}{|W|}\int\prod_{i=1}^{\text{rank}\,G}\frac{d\sigma_{i}}{\sqrt{\omega_{1}\omega_{2}}}\, & e^{-\frac{8\pi^{2}}{\omega_{1}\omega_{2}g_{YM}^{2}}\,\text{Tr}(\sigma^{2})}\cdot\left|Z_{\text{inst}}^{\left(\text{4$d$}\right)}\left(\mathfrak{q}^{\left(\text{4$d$}\right)},i\sigma,im,\omega_{1},\omega_{2}\right)\right|^{2}\\
\cdot & \prod_{\alpha\in\D_{+}}\Upsilon\left(i\alpha(\sigma)|\boldsymbol{\omega}\right)\Upsilon\left(-i\alpha(\sigma)|\boldsymbol{\omega}\right)\cdot\prod_{\rho\in\CR}\Upsilon\left(i\rho(\sigma)+im+\frac{|\boldsymbol{\omega}|}{2}\bigg|\boldsymbol{\omega}\right)^{-1}\ ,
\end{aligned}
\end{align}

\begin{itemize}
\item $g_{\text{YM}}$ is the Yang-Mills coupling. $g_{\text{YM}}$ and
$\theta_{\text{YM}}$ appear also in the instanton part of the matrix
model. $\mathfrak{q}^{\left(\text{4$d$}\right)}$ is defined in Appendix \ref{sec:The-instanton-partition}.
\item $\mathcal{R}$ is now the total representation of the hypermultiplets. 
\item We have set all hypermultiplet masses to a common value: $m$.
\item The supersymmetric R{\'e}nyi entropy is computed
using $\boldsymbol{\omega}=\left(1,\frac{1}{n}\right)$, with the
mass parameter $m$ for all hypermultiplets set to a common value
\[
m=-i\left(m_{f}^{0}-\frac{|\boldsymbol{\omega}|}{2}\right)\ .
\]

\end{itemize}

The result for a 5$d$ $\mathcal{N}=1$ theory on the squashed sphere
is still conjectural \cite{Kallen:2012cs,Hosomichi:2012ek,Imamura:2012xg,Kim:2012qf,Imamura:2012bm,Kallen:2012va,Lockhart:2012vp,Kim:2012ava}.
We follow the form in \cite{Pasquetti:2016dyl}

\begin{align}\label{5dMM}
\begin{aligned}Z^{\text{susy 5$d$}}=\frac{1}{|W|}\int&\prod_{i=1}^{\text{rank}\,G}  \left(S_{3}'(0|\bomega)\frac{d\sigma_{i}}{2\pi}\right)\,e^{\frac{1}{\omega_{1}\omega_{2}\omega_{3}}\left[-\frac{8\pi^{3}}{g_{\text{YM}}^{2}}\text{Tr}\,\sigma^{2}+i\frac{\pi}{3}\kappa\,\text{Tr}\,\sigma^{3}\right]}\\
 & \cdot\prod_{\alpha\in\D_{+}}S_{3}\left(i\alpha(\sigma)\Big|\bomega\right)S_{3}\left(-i\alpha(\sigma)\Big|\bomega\right)\prod_{\rho\in\CR}S_{3}\left(i\rho(\sigma)+im+\frac{|\bomega|}{2}\bigg|\bomega\right)^{-1}\\
 & \cdot Z_{\text{inst}}\left(\mathfrak{q},i\sigma,im,\omega_{1},\omega_{2},\frac{2\pi}{\omega_{3}}\right)Z_{\text{inst}}\left(\mathfrak{q},i\sigma,im,\omega_{3},\omega_{1},\frac{2\pi}{\omega_{2}}\right)Z_{\text{inst}}\left(\mathfrak{q},i\sigma,im,\omega_{2},\omega_{3},\frac{2\pi}{\omega_{1}}\right),
\end{aligned}
\end{align}

\begin{itemize}
\item $\kappa$ is the Chern-Simons coupling. $\kappa$ and $g_{\text{YM}}$
appear also in the instanton part. $\mathfrak{q}$ is defined in Appendix \ref{sec:The-instanton-partition}.
\item
The derivative of the triple sine function in the integral measure
can be written into \cite{kurokawa2004derivatives} 
\begin{align}
S_{3}'(0|\bomega)=\frac{\rho_{3}(\omega_{1},\omega_{2},\omega_{3})^{2}\rho_{1}(\omega_{1})\rho_{1}(\omega_{2})\rho_{1}(\omega_{3})}{\rho_{2}(\omega_{1},\omega_{2})\rho_{2}(\omega_{2},\omega_{3})\rho_{2}(\omega_{3},\omega_{1})}\ ,
\end{align}
where $\rho_{r}(\bomega)$ is the Stirling modular form 
\begin{align}
\rho_{r}(\bomega):=\lim_{z\to0}\frac{1}{z\,\Gamma_{r}(z|\bomega)}\ .
\end{align}
Note that $\rho_{1}(\omega)=\sqrt{2\pi/\omega}$.
\item $\mathbf{\omega}=\left(\omega_{1},\omega_{2},\omega_{3}\right)$ are
squashing parameters, and $\left|\mathbf{\boldsymbol{\omega}}\right|=\left|\omega_{1}+\omega_{2}+\omega_{3}\right|$.
The supersymmetric R{\'e}nyi entropy is computed
using $\boldsymbol{\omega}=\left(1,1,\frac{1}{n}\right)$, and setting
\[
m=-i\left(m_{f}^{0}-\frac{|\boldsymbol{\omega}|}{2}\right)\ .
\]
\end{itemize}

\subsection{The scalar moduli space and classical contributions}

All of the theories we consider have a moduli space which is partially
given by the vev of a Lie algebra valued scalar. The scalar $\sigma$
has eigenvalues $\vec{\sigma}$. Integration over $\sigma$ is what
makes the result of the localization procedure into a matrix model.
In our setup for the $n$-copy theory, there is one such $\sigma_{k}$
and one integration for each copy. There are also classical contributions
to the matrix model which depend on $\sigma$.

The mode which $\sigma$ parametrizes is a part of the original theory
which cannot be treated as free, even after localization. Therefore,
the values of $\sigma_{k}$ are subject to the boundary conditions
implied by the original codimension-one definition of the replica
defect. It is trivial to see that this implies that all $\sigma_{k}$
are equal.\footnote{This only makes sense if one identifies the gauge transformations
at the interface. This also implies the identification of the residual
gauge transformations acting in the different matrix models.}
Equivalently, the sewing operation for this set of moduli consists
of a set of delta functions, in the matrix model for the $n$-copy
theory in the presence of the defect, which enforce this equality
\[
\delta_{\text{moduli}}^{\text{scalar}}=\prod_{k=0}^{n-2}\prod_{i=1}^{\text{rank}\,G}\delta\left(\left(\sigma_{k}\right)_{i}-\left(\sigma_{k+1}\right)_{i}\right)\ .
\]

The classical contributions depending only on $\sigma$ are products
of expressions of the form
\[
e^{c\left(\prod_{i}\omega_{i}^{-1}\right)\text{Tr}\left(\sigma^{p}\right)}\ ,
\]
for some constant $c$. In the theory considered on the squashed sphere,
we have 
\[
\prod\omega_{i}^{-1}=n,
\]
while for the round sphere
\[
\prod\omega_{i}^{-1}=1.
\]
Starting from the $n$-copy theory on the round sphere, after using
the delta functions to set all the $\sigma_{k}$ equal, we recover
the factor of $n$.

\subsection{\label{sub:Perturbative_contributions}Perturbative contributions}

The defect operator interpretation of the supersymmetric R{\'e}nyi
entropy was originally observed in \cite{Nishioka:2013haa} by rewriting
the perturbative partition function on the $n$-fold cover as $n$-copies of the
partition functions on a round three-sphere with vortex loops inserted
on each copy. (See also \cite{Giveon:2015cgs,Mori:2015bro} for related
works in two dimensions). We extend this interpretation to higher
dimensions and show the perturbative parts of the partition functions
in the 4$d$ and $5d$ supersymmetric R{\'e}nyi entropies
also have similar structures. The non-perturbative contributions arise
in higher dimensions will be discussed separately in Section \ref{ss:nonperturvative}.

First, let us review the story in three dimensions \cite{Nishioka:2013haa}.
There are no non-perturbative contributions in the matrix model \eqref{3dMM} and we only need to deal with the one-loop partition functions appearing as the double sine functions $S_{2}$ from the vector and matter multiplets.
The identity (\ref{Replica_Identity}) for the double sine function yields
that the one-loop partition function of a multiplet with R-charge $\Delta_I$ can be decomposed as a
product of those in the presence of a supersymmetric abelian vortex
loop \cite{Drukker:2012sr} 
\begin{align}
\begin{aligned}S_{2}\left(i\rho(\sigma)+\frac{|\bomega|}{2}\D_{I}\bigg|1,\frac{1}{n}\right)^{-1} & =\prod_{k=0}^{n-1}S_{2}\left(i\rho(\sigma)+\frac{\Delta_{I}}{2}\left(1+\frac{1}{n}\right)+\frac{k}{n}\bigg|1,1\right)^{-1}\ ,\\
 & =\prod_{k=0}^{n-1}S_{2}\left(i\rho(\sigma)+q_{k}^{3d~\text{vortex}}+\Delta_{I}\bigg|1,1\right)^{-1}\ ,
\end{aligned}
\end{align}
where $q_{k}^{3d~\text{vortex}}$ is introduced to be 
\begin{align}\label{SurfaceCharge3d}
q_{k}^{3d~\text{vortex}}=\frac{\Delta_{I}}{2}\left(\frac{1}{n}-1\right)+\frac{k}{n}\ .
\end{align}
Comparing it with the shift of the modulus $\sigma$ (\ref{3d_defect_shift}),
we interpret the decomposition as a manifestation of the introduction
of a supersymmetric abelian vortex loop, described in Section \ref{sub:3d_defect},
of charge $q_{k}^{\text{vortex}}$ supported on the entangling surface
$(\theta=0)$ on each copy of a round sphere. Note that $q_{k}^{\text{vortex}}$
differs from (\ref{3d_defect_shift}) by a term proportional to the
R-charge $\Delta_{I}$ for matter multiplets. This means that the vortex loops
for the supersymmetric R{\'e}nyi entropy are dressed
by the R-symmetry flux.

One can work out a similar decomposition for the perturbative part in the 4$d$ $\CN=2$ matrix model \eqref{4dMM} with a slight modification.
The matter one-loop partition function is represented by the $\Upsilon$ function, which enjoys decomposition into the product of the $n$-copies with the help of the identity (\ref{4dSurfaceFormula}): 
\begin{align}
\Upsilon\left(i\rho(a)+im+\frac{|\bomega|}{2}\bigg|1,\frac{1}{n}\right)^{-1}=\prod_{k=0}^{n-1}\Upsilon\left(i\rho(a)+1+q_k^\text{vortex}\bigg|1,1\right)^{-1}\ ,
\end{align}
where $q_k^\text{vortex}$ is to be interpreted as the charge
of a codimension-two surface defect 
\begin{align}
q_k^\text{vortex}=\frac{k}{n}\ ,\label{SurfaceCharge}
\end{align}
under the choice of the shifted mass \eqref{4d_mass_shift} and the relation \eqref{mass_relation}.
It agrees with the shift of the modulus \eqref{4dsurfacecharge} induced by the insertion of a supersymmetric abelian surface operator of charge
$k$ in Section \ref{sub:4d_defect}.

Repeating the analogous procedure to the 5$d$ $\CN =1$ theory one finds the matter one-loop partition function with the identity \eqref{Replica_Identity}:
\begin{align}
\begin{aligned}S_{3}\left(i\rho(\sigma)+im+\frac{|\bomega|}{2}\bigg|1,1,\frac{1}{n}\right)^{-1} & =\prod_{k=0}^{n-1}S_{3}\left(i\rho(\sigma)+1 + q_k^\text{vortex}\bigg|1,1,1\right)^{-1}\ ,\end{aligned}
\end{align}
where we introduce the charges $q_k^\text{vortex}$ by \eqref{SurfaceCharge} and the mass shift \eqref{5d_mass_shift_new}.
Once again it can be interpreted as a supersymmetric codimension-two
surface defect of charge $k$ described in Section \ref{sub:5d_defect}.

It is straightforward to apply the same argument to vector multiplets in any dimensions to read off the surface charges and the results \eqref{SurfaceCharge3d} and \eqref{SurfaceCharge} still hold with $\Delta_I = 0$.

\subsection{Non-perturbative contributions - Instantons and contact-instantons}\label{ss:nonperturvative}

The partition functions on the four-sphere and the five-sphere receive
non-perturbative contributions from instantons and contact-instantons,
respectively. These are supersymmetric configurations localized at
fixed loci of the equivariant action generated by the square of
the supersymmetry. For the four-sphere, the fixed points are at the
north and south pole and give rise to instanton and anti-instanton
contributions, respectively. The five-sphere partition function includes
contributions from contact instantons \cite{Kallen:2012cs}, which
are extended along the fiber of 
\[
\BS^1\rightarrow \BS^5\rightarrow\mathbb{C}P^{2}\ ,
\]
and localized at three points on the base.\footnote{To the best of our understanding, the precise form of these contributions
is still conjectural. We will use the form considered in e.g. \cite{Pasquetti:2016dyl}
and find that it works well.}
The supersymmetric instanton contributions are computed by the Nekrasov
partition function \cite{Nekrasov:2002qd,Nekrasov:2003rj}. We find
it convenient to express both types of contributions in terms of the
$5d$, or $q$-deformed, version of the Nekrasov partition function,
which we review in Appendix \ref{sec:The-instanton-partition}. The $4d$ undeformed
partition function can be recovered by taking an appropriate limit.

Instanton and contact-instanton contributions introduce new moduli,
classical contributions, and perturbative contributions into the calculation.
We begin by discussing how the moduli of the replicas in the $n$-copy
theory are sewn up to produce those of the original theory. We assert
a specific pattern for the fractionalization of a particular instanton
configuration, given by a vector of Young diagrams. We then show that,
given this pattern, the classical and perturbative contributions recombine
to yield \eqref{eq:defect_calculation}. 

Similar decompositions of partitions appear in the context of instantons
on ALE spaces (c.f.\,\cite{Fucito:2004ry}). A relationship between
the instanton partition function on a $\mathbb{Z}_{n}$ orbifold of
$\mathbb{C}^{2}$, acting on just one $\mathbb{C}$ factor, and surface
operators appears in \cite{Kanno:2011fw,Wyllard:2010vi} (see also
the review \cite{Tachikawa:2014dja}). It is possible that the results
we need for the covering space associated to the $n$-copy theory
can be recovered from the latter papers. Specifically, this seems
plausible given the connection between the supersymmetric R{\'e}nyi entropy
at $n$ and at $1/n$. However, we need $5d$ results and a very specific
surface operator, so we derive the necessary fractionalization and
relationships between the determinants in this context.

\subsubsection{Sewing of instantons}

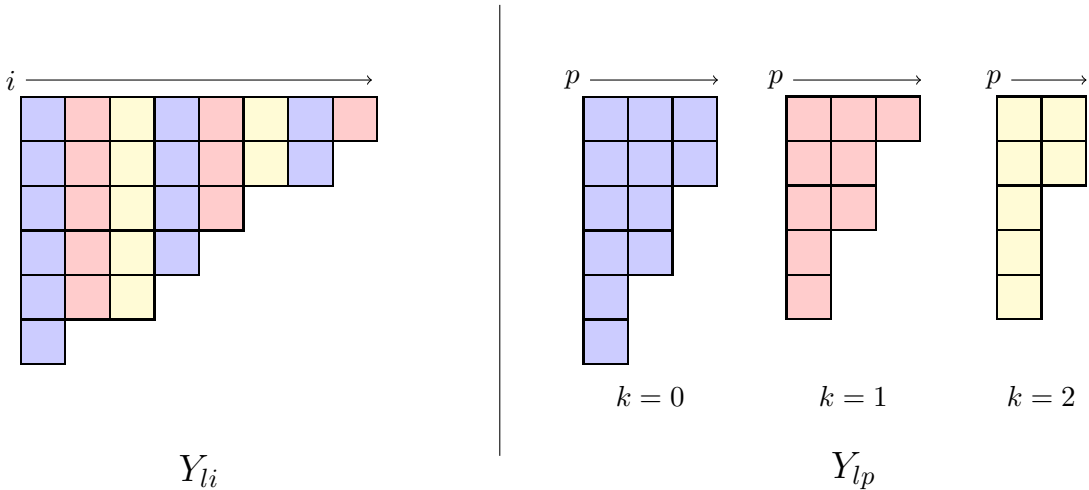
\begin{figure}[htbp]
\centering
\begin{tikzpicture}
	\node at (0,0) {
	\ytableaushort{\none}
 *{8,7,5,4,3,,1}
 *[*(blue!20)]{1,1,1,1,1,1}
 *[*(red!20)]{2,2,2,2,2}
 *[*(yellow!20)]{3,3,3,3,3}
 *[*(blue!20)]{4,4,4,4}
 *[*(red!20)]{5,5,5}
 *[*(yellow!20)]{6,6}
 *[*(blue!20)]{7,7}
 *[*(red!20)]{8}
	};
	\draw[->] (-2.3,2) node[left] {$i$} -- (2.3,2);
	\node at (0,-3.2) {\Large $Y_{li}$};
	
	\draw[-] (4,3) -- (4,-3);
	
	\node at (6,0) {
	\ytableaushort{\none}
 *[*(blue!20)]{3,3,2,2,1,1}
	};
	\draw[->] (5.2,2) node[left] {$p$} -- (6.9,2);
	\node at (6,-2.2) {$k=0$};
	
	\node at (8.7,0.3) {
	\ytableaushort{\none}
 *[*(red!20)]{3,2,2,1,1}
	};
	\draw[->] (7.9,2) node[left] {$p$} -- (9.6,2);
	\node at (8.7,-2.2) {$k=1$};
	
	\node at (11.2,0.3) {
	\ytableaushort{\none}
 *[*(yellow!20)]{2,2,1,1,1}
	};
	\draw[->] (10.8,2) node[left] {$p$} -- (11.8,2);
	\node at (11.2,-2.2) {$k=2$};
	
	\node at (8.7,-3.2) {\Large $Y_{lp}$};
\end{tikzpicture}
\caption{Fractionalization and recombination
of the Young diagrams representing singular instantons for $n=3$.}
\label{fig:Fractionalization}
\end{figure}

After employing equivariant localization, the instanton moduli space
localizes to a set of points, each given by a vector of partitions
\cite{Nekrasov:2002qd,Nekrasov:2003rj}. The partitions describing
the instanton moduli space of the theory on the branched sphere fractionalize
to yield partitions for each of the replicas in the $n$-copy theory.
This happens by splitting one set of Young diagrams, yielding a partition
vector $\vec{\mathbf{Y}}$, into $n$ Young diagrams, yielding partitions
$\left\{ \vec{\mathbf{Y}}\left(k\right)\right\} _{k=0}^{n-1}$, having
the same combined number of boxes. This process is illustrated in
figure \ref{fig:Fractionalization}. The diagrams can be split along
the vertical or along the horizontal, depending on which deformation
parameter $\epsilon_{i}$ is taken to be $n$ times smaller. Since
the instanton partition function is invariant under the simultaneous
transposition of the diagrams and $\epsilon_{1}\leftrightarrow\epsilon_{2}$,
it suffices to consider the situation in figure \ref{fig:Fractionalization}.

For the contribution of a hypermultiplet in the fundamental representation,
each element of the fluctuation determinant around an instanton can
be associated to a box in the Young diagram, and figure \ref{fig:Fractionalization}
describes the decomposition into replicas. The contribution from a
vector multiplet, or from an adjoint hypermultiplet, involves pairs
of partitions in the vector $\vec{\mathbf{Y}}$. We find that the
decomposition into replicas follows the pattern
\[
\left(\mathbf{Y}_{1},\mathbf{Y}_{2}\right)~\rightarrow~\prod_{k_{1,2}=0}^{n-1}\left(\mathbf{Y}_{1}\left(k_{1}\right),\mathbf{Y}_{2}\left(k_{2}\right)\right)|_{k_{1}-k_{2}=k\text{ mod }n}\ .
\]
The moduli space of contact-instantons can also fractionalize along
the additional $\BS^1$ direction. This process is simpler, amounting
to decomposing the Kaluza-Klein (KK) momentum, and is described in Section \ref{sub:The-third-point}.
It does not involve the partitions.

\subsubsection{Classical and Chern-Simons contributions depending on instantons}

The classical weight in the instanton partition function of a configuration
with instanton number $|\vec{{\bf Y}}|$ in a 5$d$ SCFT is
\[
\mathfrak{q}^{|\vec{{\bf Y}}|}\ .
\]
The number of boxes in the Young diagrams on the right pane of figure \ref{fig:Fractionalization}
sum to the number of boxes on the left, and therefore the combined
weight from each of the replicas matches that of the original theory
\[
\mathfrak{q}^{|\vec{{\bf Y}}|}=\prod_{k=0}^{n-1}\mathfrak{q}^{|\vec{{\bf Y}}\left(k\right)|}\ .
\]

In the presence of a $5d$ Chern-Simons term, we also need to split
the contribution
\[
Z_{\vec{{\bf Y}},\kappa}^{\text{5$d$-CS}}\left({\vec{a}},\epsilon_{1},\epsilon_{2},\beta\right)=\exp\left[i\beta\,\kappa\sum_{l}\sum_{\left(s,t\right)\in\mathbf{Y}_{l}}\left(a_{l}-\left(s-1\right)\epsilon_{1}-\left(t-1\right)\epsilon_{2}\right)\right].
\]
We do this first for the case where $\epsilon_{1}=1/n$ by setting
$s=np-k$ 
\begin{align}
\begin{aligned}
Z_{\vec{{\bf Y}},\kappa}^{\text{5$d$-CS}}\left({\vec{a}},\frac{1}{n},1,2\pi\right) & =\exp\left[2\pi i\,\kappa\sum_{l}\sum_{\left(s,t\right)\in\mathbf{Y}_{l}}\left(a_{l}-\frac{s-1}{n}-\left(t-1\right)\right)\right]\ ,\\
 & =\prod_{k=0}^{n-1}\exp\left[2\pi i\,\kappa\sum_{l}\sum_{\left(p,t\right)\in\mathbf{Y}_{l}\left(k\right)}\left(a_{l}-\frac{k}{n}-\left(p-1\right)-\left(t-1\right)\right)\right]\ ,
\end{aligned}
\end{align}
where in the second line we have reparametrized 
\[
k~\rightarrow~ n-1-k\ .
\]
The expression on the second line can be thought of as the contributions
from the $n$-replicas, where the additional shift 
\[
a_{l}~\rightarrow~ a_{l}-\frac{k}{n} \ ,
\]
is the effect of the monodromy brought on by the defect. Since we
associate this shift with 
\[
q_k^\text{surface}=\frac{k}{n}\ ,
\]
we see that the contribution from the Higgs type mass enters the Chern-Simons
term in the same way as it enters the fluctuation determinant for
the fundamental hypermultiplet considered below.

The contribution from the $k^{th}$ replica is sensitive only to the
partition represented by the Young diagrams $\vec{{\bf Y}}\left(k\right)$.
It might seem strange to see the mass shift appear at all in a classical
contribution. Note, however, that the combination appearing in the
exponential comes from evaluating the classical Chern-Simons term
at the positions of the poles for the integral over the scalar associated
to the auxiliary $U\left(|\vec{{\bf Y}}|\right)$ symmetry \cite{Kim:2008kn,Collie:2008vc,Kim:2012gu}.
This position is shifted by the Higgs type mass term.

\subsubsection{Fundamental hypermultiplets fluctuations }

We now demonstrate the relationship between the contribution of a
hypermultiplet on the squashed sphere, with deformation parameters
$\left(\epsilon_{1},\epsilon_{2},\beta\right)=\left(\frac{1}{n},1,2\pi\right)$,
and $n$ hypermultiplets in the presence of defects with deformation
parameters $\left(\epsilon_{1},\epsilon_{2},\beta\right)=\left(1,1,2\pi\right)$.
At the conformal point 
\[
Q_{m_{f}^{0}}^{-1}=t\ .
\]
Expressing the squashed sphere contribution using the round sphere
values for $t,q$ etc., we get 
\begin{align}
\begin{aligned}
Z_{\vec{{\bf Y}}}^{\text{fund hyper}}\left(\vec{a},m_{f}^{0},\frac{1}{n},1,2\pi\right) & =\prod_{l=1}^{N}\prod_{j=1}^{\infty}\frac{\left(Q_{l}\,q\,t^{-\frac{j-1}{n}};q\right)_{\infty}}{\left(Q_{l}\,q^{Y_{lj}+1}\,t^{-\frac{j-1}{n}};q\right)_{\infty}}\ ,\\
 & =\prod_{k=0}^{n-1}\left[\prod_{l=1}^{N}\prod_{j=1}^{\infty}\frac{\left(Q_{l}\,q\,t^{-\frac{j-1}{n}};q\right)_{\infty}}{\left(Q_{l}\,q^{Y_{lj}+1}\,t^{-\frac{j-1}{n}};q\right)_{\infty}}\right]{}_{j=np-k}\ ,\\
 & =\prod_{k=0}^{n-1}\left[\prod_{l=1}^{N}\prod_{p=1}^{\infty}\frac{\left(Q_{m_{f}}^{-1}\left(k\right)Q_{l}\,q\,t^{-\left(p-1\right)};q\right)_{\infty}}{\left(Q_{m_{f}}^{-1}\left(k\right)Q_{l}\,q^{Y_{l,np-k}+1}\,t^{-\left(p-1\right)};q\right)_{\infty}}\right]\ ,
\end{aligned}
\end{align}
and 
\[
Q_{m_{f}}^{-1}\left(k\right)\equiv t^{\frac{k}{n}-1+\frac{1}{n}}\ .
\]
Reparametrizing the product over $k$ as $k\rightarrow n-1-k$,
we get
\[
Q_{m_{f}}^{-1}\left(k\right)\equiv t^{-\frac{k}{n}}\ ,
\]
implying
\[
m_{f}=\frac{k}{n}\ .
\]

We now identify the terms in the square parentheses with the contribution
of fluctuations of a hypermultiplet in the presence of a codimension-two defect
\begin{align}
Z_{\vec{\mathbf{Y}}}^{\text{hyper-defect}}\left(\vec{a},1,1,2\pi;k\right)\equiv\left[\prod_{l=1}^{N}\prod_{p=1}^{\infty}\frac{\left(Q_{m_{f}}^{-1}\left(k\right)Q_{l}\,q\,t^{-\left(p+1\right)};q\right)_{\infty}}{\left(Q_{m_{f}}^{-1}\left(k\right)Q_{l}\,q^{Y_{l,np+k}+1}\,t^{-\left(p+1\right)};q\right)_{\infty}}\right]\ ,
\end{align}
The expression for $Z_{\vec{\mathbf{Y}}}^{\text{hyper-defect}}\left(k\right)$
differs from the expression $Z_{\vec{\mathbf{Y}}}^{\text{hyper}}$
in two ways
\begin{enumerate}
\item The fugacity, or mass parameter, involving the background vector is
shifted in the imaginary direction by 
\[
q_{\text{vortex}}=\frac{k}{n}\ .
\]
We ascribe this to the effect of the codimension-two defect on the
fluctuations.
\item The $k^{th}$ such contribution is sensitive only to the boxes of the
Young diagram with horizontal position given by 
\[
i=k\text{ mod }n\ .
\]
We ascribe this to the fractionalization of the instanton moduli corresponding
to the partition. 
\end{enumerate}
These are the same effects visible for the Chern-Simons contribution.
We conclude that
\[
Z_{\vec{{\bf Y}}}^{\text{fund hyper}}\left(\vec{a},m_{f}^{0},\frac{1}{n},1,2\pi\right)=\prod_{k=0}^{n-1}Z_{\vec{\mathbf{Y}}}^{\text{hyper-defect}}\left(\vec{a},1,1,2\pi;k\right)\ .
\]

\subsubsection{Adjoint hypermultiplet or vector multiplet fluctuations}

An adjoint hypermultiplet contributes to the fluctuation determinant
around an instanton configuration as 
\begin{align}
Z_{\vec{{\bf Y}}}^{\text{adjoint hyper}}\left(\vec{a},m_{f}^{0},\epsilon_{1},\epsilon_{2},\beta\right)=\prod_{\left(l,i\right)\ne\left(m,j\right)}\frac{\left(Q_{m_{f}^{0}}\,Q_{l}\,Q_{m}^{-1}\,q^{Y_{li}-Y_{mj}}\,t^{j-i+1};q\right)_{\infty}}{\left(Q_{m_{f}^{0}}^{-1}\,Q_{l}\,Q_{m}^{-1}\,q^{Y_{li}-Y_{mj}}\,t^{j-i};q\right)_{\infty}}\frac{\left(Q_{m_{f}^{0}}^{-1}\,Q_{l}\,Q_{m}^{-1}\,t^{j-i};q\right)_{\infty}}{\left(Q_{m_{f}^{0}}\,Q_{l}\,Q_{m}^{-1}\,t^{j-i+1};q\right)_{\infty}}\ .
\end{align}
This involves pairs of partitions $Y_{l}$ and $Y_{m}$. The virtue
of the form of the fluctuation determinant written above is that $t$
appears raised only to a power corresponding the the column indices
$i,j$. As such, the determinant can be decomposed in a way similar
to the fundamental hypermultiplet
\begin{align}
\begin{aligned}
Z_{\vec{{\bf Y}}}^{\text{adjoint hyper}}&\left(\vec{a},\frac{1}{n},1,1,2\pi\right) \\
 & =\prod_{\left(l,i\right),\left(m,j\right)}\frac{\left(Q_{l}\,Q_{m}^{-1}\,q^{Y_{li}-Y_{mj}}\,t^{\frac{1}{n}\left(j-i\right)};q\right)_{\infty}}{\left(Q_{l}\,Q_{m}^{-1}\,q^{Y_{li}-Y_{mj}}\,t^{\frac{1}{n}\left(j-i+1\right)};q\right)_{\infty}}\frac{\left(Q_{l}\,Q_{m}^{-1}\,t^{\frac{1}{n}\left(j-i+1\right)};q\right)_{\infty}}{\left(Q_{l}\,Q_{m}^{-1}\,t^{\frac{1}{n}\left(j-i\right)};q\right)_{\infty}}\ ,\\
 & =\prod_{k=0}^{n-1}\left[\prod_{\left(l,i\right),\left(m,j\right)}^{j-i=n\left(p-q\right)+k}\frac{\left(Q_{l}\,Q_{m}^{-1}\,q^{Y_{li}-Y_{mj}}\,t^{p-q+\frac{k}{n}};q\right)_{\infty}}{\left(Q_{l}\,Q_{m}^{-1}\,q^{Y_{mj}-Y_{li}}\,t^{p-q+1-\frac{k}{n}};q\right)_{\infty}}\frac{\left(Q_{l}\,Q_{m}^{-1}\,t^{p-q+1-\frac{k}{n}};q\right)_{\infty}}{\left(Q_{l}\,Q_{m}^{-1}\,t^{p-q+\frac{k}{n}};q\right)_{\infty}}\right]\ ,\\
 & =\prod_{k=0}^{n-1}\left[\prod_{\left(l,i\right),\left(m,j\right)}^{i-j=n\left(p-q\right)+k}\frac{\left(Q_{m_{f}}\left(k\right)\,Q_{l}\,Q_{m}^{-1}\,q^{Y_{li}-Y_{mj}}\,t^{p-q};q\right)_{\infty}}{\left(Q_{m_{f}}^{-1}\left(k\right)\,Q_{l}\,Q_{m}^{-1}\,q^{Y_{mj}-Y_{li}}\,t^{p-q+1};q\right)_{\infty}}\frac{\left(Q_{m_{f}}^{-1}\left(k\right)\,Q_{l}\,Q_{m}^{-1}\,t^{p-q+1};q\right)_{\infty}}{\left(Q_{m_{f}}\left(k\right)\,Q_{l}\,Q_{m}^{-1}\,t^{p-q};q\right)_{\infty}}\right]\ .
\end{aligned}
\end{align}
In the second line, we have replaced in two of the factors $k\rightarrow n-1-k$.
If we define
\begin{align}
\begin{aligned}
Z_{\vec{\mathbf{Y}}}^{\text{adjoint -defect}}&\left(\vec{a},1,1,2\pi;k\right)\\&\equiv\prod_{\left(l,i\right),\left(m,j\right)}^{i-j=n\left(p-q\right)+k}\frac{\left(Q_{m_{f}}\left(k\right)\,Q_{l}\,Q_{m}^{-1}\,q^{Y_{li}-Y_{mj}}\,t^{p-q};q\right)_{\infty}}{\left(Q_{m_{f}}^{-1}\left(k\right)\,Q_{l}\,Q_{m}^{-1}q^{Y_{mj}-Y_{li}}\,t^{p-q+1};q\right)_{\infty}}\frac{\left(Q_{m_{f}}^{-1}\left(k\right)\,Q_{l}\,Q_{m}^{-1}\,t^{p-q+1};q\right)_{\infty}}{\left(Q_{m_{f}}\left(k\right)\,Q_{l}\,Q_{m}^{-1}\,t^{p-q};q\right)_{\infty}}\ ,
\end{aligned}
\end{align}
such that
\[
Z_{\vec{{\bf Y}}}^{\text{adjoint hyper}}\left(\vec{a},m_{f}^{0},\frac{1}{n},1,2\pi\right)=\prod_{k=0}^{n-1}Z_{\vec{\mathbf{Y}}}^{\text{adjoint-defect}}\left(\vec{a},1,1,2\pi;k\right)\ .
\]

We now consider the contribution of a vector multiplet. Its fluctuation
determinant is inverse to that of an adjoint hypermultiplet with zero
mass, as was the case for the perturbative contribution. As we did
there, we keep a ``mass fugacity'' to keep track of the deformation
brought on by the defect. Since this mass is now associated to the
Higgs vev, it is not shifted from $m_{f}=0$ by the Okuda-Pestun prescription.
Repeating the calculation for the adjoint hypermultiplet using the
same manipulations, we get 
\[
Z_{\vec{{\bf Y}}}^{\text{vector}}\left(\vec{a},m_{f}^{0},\frac{1}{n},1,2\pi\right)=\prod_{k=0}^{n-1}Z_{\vec{\mathbf{Y}}}^{\text{vector-defect}}\left(\vec{a},1,1,2\pi;k\right)\ ,
\]
with
\begin{align}
	\begin{aligned}
Z_{\vec{\mathbf{Y}}}^{\text{vector-defect}}&\left(\vec{a},1,1,2\pi;k\right)\\
	&\equiv\prod_{\left(l,i\right)\ne\left(m,j\right)}^{i-j=n\left(p-q\right)+k}\frac{\left(Q_{m_{f}}\left(k\right)\,Q_{l}\,Q_{m}^{-1}\,q^{Y_{li}-Y_{mj}}\,t^{p-q};q\right)_{\infty}}{\left(Q_{m_{f}}^{-1}\left(k\right)\,Q_{l}\,Q_{m}^{-1}\,q^{Y_{mj}-Y_{li}}\,t^{p-q+1};q\right)_{\infty}}\frac{\left(Q_{m_{f}}^{-1}\left(k\right)\,Q_{l}\,Q_{m}^{-1}\,t^{p-q+1};q\right)_{\infty}}{\left(Q_{m_{f}}\left(k\right)\,Q_{l}\,Q_{m}^{-1}\,t^{p-q};q\right)_{\infty}}\ .
	\end{aligned}
\end{align}

Since the original product is over pairs of Young diagrams corresponding
to $l$ and $m$, the decomposition of an adjoint hypermultiplet or
a vector multiplet into $n$ parts is not as simple as in figure \ref{fig:Fractionalization}.
Instead, each pair of diagrams of the theory on the squashed sphere
splits into $n$ pairs for each of the $n$ copies.

\subsubsection{\label{sub:The-third-point}The third point}

The $5d$ squashed sphere has one more contribution, not of the type
above. The third point contributes a sum over contact-instantons with
deformation parameters $\left(\epsilon_{1},\epsilon_{2},\beta\right)=\left(1,1,2\pi n\right)$,
i.e. $\beta$ is $n$ times as large as it would be on the round sphere.
In order to decompose this contribution, it is useful to write the
determinant part of the $q$-deformed instanton partition function as
a product over Kaluza Klein modes coming from the extra circle. Starting
from the expressions in \cite{Awata:2008ed}
\[
Z_{\vec{{\bf Y}}}^{\text{vector}}\left(\vec{a},\epsilon_{1},\epsilon_{2},\beta\right)=\prod_{l,m}^{N_{c}}\left(N_{l,m}^{\vec{{\bf Y}}}\right)^{-1},
\]
where 
\begin{align}
N_{l,m}^{\vec{{\bf Y}}}=\prod_{s\in Y_{l}}\left[1-e^{i\beta\left(\ell_{Y_{m}}\left(s\right)\epsilon_{1}-(a_{Y_{l}}\left(s\right)+1)\epsilon_{2}+a_{l}-a_{m}\right)}\right] \prod_{t\in Y_{m}}\left[1-e^{i\beta\left(-(\ell_{Y_{l}}\left(t\right)+1)\epsilon_{1}+a_{Y_{m}}\left(t\right)\epsilon_{2}+a_{l}-a_{m}\right)}\right]\ ,
\end{align}
the defect decomposition, in this case, follows simply from the identity
\[
\prod_{k=0}^{n-1}\left[1-e^{2\pi i\left(\alpha+\frac{k}{n}\right)}\right] =1-e^{2\pi in\alpha}\ .
\]
A similar expression exists for the hypermultiplet contribution. Its
decomposition follows from the same method.\footnote{The mass shift given by $m_{f}^{0}$ is immaterial in this case.}

To relate this decomposition to the KK decomposition, we use the regularized
infinite product
\[
\prod_{m=-\infty}^{\infty}\left(m+a\right)=1-e^{2\pi ia}\ ,\qquad\text{Im}\left(a\right)>0\ .
\]
The defect decomposition can now be thought of as writing the quantum
number $m$ in the form 
\[
m=np+k\ .
\]
The partitions of the various copies are simply identified in this
case, in analogy with the vev of the scalar modulus.

\subsection{$n\rightarrow1/n$ duality}

The supersymmetric R{\'e}nyi entropy in three and four dimensions satisfies
an interesting property stemming from the fact that for a superconformal
theory
\begin{equation}
Z_{n}^{\text{3$d$ or 4$d$}}=Z_{1/n}^{\text{3$d$ or 4$d$}}\ .\label{eq:n_inversion}
\end{equation}
This follows simply from two facts:
\begin{enumerate}
\item Conformal invariance implies a dependence on the squashing parameters
$\omega_{1/2}$ of the form
\[
Z^{\text{3$d$ or 4$d$}}\left(\frac{\omega_{1}}{\omega_{2}}\right)\ .
\]

\item There is a trivial change of coordinates which exchanges 
\[
\omega_{1}~\leftrightarrow~\omega_{2}\ .
\]
\end{enumerate}
Taking $\omega_{1}=1/n$ and $\omega_{2}=1$ yields \eqref{eq:n_inversion}.
The same trick does not work in five dimensions. This relationship
can be thought of as an interacting supersymmetric version of the
Bose-Fermi duality in three dimensions \cite{Bueno:2015qya}, which does not hold for R{\'e}nyi entropies in higher dimensions without introducing supersymmetry.

One can calculate the $n^{th}$ supersymmetric R{\'e}nyi entropy from the
partition function on the branched sphere. From the point of view
of the instanton partition function, taking 
\[
\epsilon_{1}=1/n\ ,\qquad\epsilon_{2}=1\ ,
\]
corresponds to counting instantons on a space which is branched over
a codimension-two surface. Taking 
\[
\epsilon_{1}=n\ ,\qquad\epsilon_{2}=1\ ,
\]
on the other hand, corresponds to counting instantons on an orbifold.
It is interesting that the two counts are related.

\section{Discussion}

We have shown that the supersymmetric R{\'e}nyi entropy
(SRE) can be computed using supersymmetric codimension-two defects.
After giving a microscopic definition of the defect operators, we
computed the expectation values of these defects using localization.\footnote{The supersymmetric codimension-two defects coincide, in three and
in four dimensions, with specific versions of the operators defined
in \cite{Kapustin:2012iw} and \cite{Gukov:2006jk} respectively.
Five-dimensional versions were considered in e.g. \cite{Bullimore:2014upa}.} We showed that the effect of such defects on the matrix models calculating
the partition function on the round sphere amounted to imaginary mass
terms. We made a conjecture regarding the details of the sewing operation
needed to complete the picture for the moduli, scalar vevs, instantons
and contact-instantons, encountered in localization. We then showed
the equality with the squashed sphere partition function.  

Although we explicitly only showed agreement of the partition functions
representing the SRE, the decomposition into defects seems to work
at the level of the matrix model ingredients, and for any deformation
parameters $\omega_{i}$. It is reasonable to conjecture that it works
at the level of the 5$d$ holomorphic blocks and gluing \cite{Nieri:2013vba,Nieri:2013yra,Pasquetti:2016dyl}.
If this is the case, a relationship similar to the one described here
should hold for the partition functions on four-manifolds and five-manifolds
of the type described in e.g. \cite{Festuccia:2016gul,Qiu:2015rwp}. 

In the context of holographic duality, the Ryu-Takanayagi prescription \cite{Ryu:2006bv,Ryu:2006ef}
allows us to compute the entanglement entropy in a CFT, in a particular
limit corresponding to classical gravity in the bulk, using a minimal
area surface in AdS which is homologous to a given entangling region $\Sigma$. 
Corrections to this computation have recently been conjectured in
\cite{Faulkner:2013ana,Barrella:2013wja}. A variant for the R{\'e}nyi
entropy was put forth in \cite{Dong:2016fnf}. Somewhat similar prescriptions
are used to compute the expectation values of supersymmetric non-local
operators (see e.g. \cite{Maldacena:1998im,Rey:1998ik,Drukker:1999zq,Constable:2002xt,Gomis:2007fi,Drukker:2008wr,Koh:2008kt}).
The authors, and others, have long suspected that there
is a relationship between these computations. We do not, however,
know of a concrete example of such a relationship. We hope that the
definition of the supersymmetric defect operator version of the SRE
calculation can be used to find one. This may involve going, first,
to a dual picture in the SCFT. For instance, the codimension-two
defects realizing the SRE in a 3$d$ $\mathcal{N}=2$ theory are vortex
loops, which, in certain situations, are dual to a Wilson loop under 3$d$ mirror
symmetry \cite{Kapustin:2012iw}.

\acknowledgments 
We would like to thank S.\,Hellerman, C.\,Herzog, K.\,Hosomichi, D.\,Jafferis, R.\,Myers, T.\,Okuda, A.\,Sheshmani, Y.\,Tachikawa and B.\,Willett for valuable discussions.
The work of T.\,N. was supported in part by JSPS Grant-in-Aid for Young Scientists (B) No.\,15K17628 and JSPS Grant-in-Aid for Scientific Research (A) No.16H02182. The work of I.\,Y. was supported by World Premier International Research Center Initiative (WPI), MEXT, Japan.


\appendix

\section{\label{sec:Conventions}Conventions}

We summarize our conventions for gauge theories and the matrix models
resulting from the localization procedure in three, four, and five dimensions.
To begin with, we set an overall scale associated with the size of $\mathbb{S}^{3,4,5}$
\[
\ell =1\ .
\]
Dimensionful parameters such as $\epsilon_{1,2},\omega_{1,2,3},a,m,\beta$
etc. are expressed using this scale. 

We use physics conventions for the gauge and flavor symmetry groups.
The generators of the Lie algebra $u\left(N\right)$ are taken to
be Hermitian matrices, and factors of $i$ appear in appropriate places
in the field strength. Consequently, integration over the Cartan sub-algebra
means an $N$-dimensional real integral over variables denoted $\vec{\sigma}$,
which are the eigenvalues of a matrix $\sigma$. In 5$d$, $\vec{\sigma}$
is related to the scalar vev as 
\[
\left\langle \phi\right\rangle =i\,\sigma\ .
\]
In 4$d$ we have 
\[
\left\langle \phi\right\rangle =\left\langle \bar{\phi}\right\rangle =-i\,\frac{\sigma}{2}\ .
\]
In 3$d$, where the real adjoint scalar in the vector multiplet is also
denoted $\sigma$, we have 
\[
\left\langle \sigma\right\rangle =\sigma\ .
\]
This convention extends to mass parameters, which are vevs for scalars
in background vector multiplets. The physical mass of a chiral multiplet
or hypermultiplet is a real number $m$, which bears the same relation
to the background vev as $\sigma$ does to the dynamical vev. The
deformation parameters $a$ and $m_{f}$, which are used when discussing
the instanton contributions to the partition function in 4$d$ and 5$d$,
are set to 
\[
a=i\,\sigma\ ,\qquad m_{f}=i\,m\ . \label{mass_relation}
\]

Our conventions for spinors and supersymmetry transformations are
different in different dimensions. However, supersymmetry transformation
parameters are always taken to be commuting spinors.

\section{\label{sec:Special-functions}Special functions}
We summarize the definitions and identities for the special functions appear in the text.

\paragraph{Multiple gamma function}

For $\bomega=(\omega_{1},\cdots,\omega_{r})\ge0$ and $z\in\BC$,
the multiple Hurwitz zeta function is defined by 
\begin{align}
\zeta_{r}(s,z,\bomega):=\sum_{\boldsymbol{n}\ge0}(\boldsymbol{n}\cdot\bomega+z)^{-s}\ ,
\end{align}
where $\boldsymbol{n}=(n_{1},\cdots,n_{r})\ge0$. The integral representation
is 
\begin{align}
\zeta_{r}(s,z,\bomega)=\frac{1}{\Gamma(s)}\int_{0}^{\infty}\frac{e^{-zt}}{\prod_{i=1}^{r}(1-e^{-\omega_{i}t})}t^{s-1}dt\ .
\end{align}

For an integer $N$, one can prove the identity 
\begin{align}
\zeta_{r}\left(s,z,\omega_{1},\cdots,\omega_{r-1},\frac{\omega_{r}}{N}\right)=\sum_{k=0}^{N-1}\zeta_{r}\left(s,z+\frac{k\omega_{r}}{N},\bomega\right)\ .\label{ZetaFormula}
\end{align}

The Barnes multiple gamma function $\Gamma_{r}(z|{\boldsymbol{\omega}})$
is defined by 
\begin{align}
\Gamma_{r}(z|\bomega):=\exp\left[\partial_{s}\zeta_{r}(s,z,\bomega)|_{s=0}\right]\ .
\end{align}

\paragraph{Multiple sine function}

One can define the $r$-ple sine function $S_{r}(z|\bomega)$ by 
\begin{align}
S_{r}(z|\bomega):=\Gamma_{r}(z|\bomega)^{-1}\Gamma_{r}(|\bomega|-z|\bomega)^{(-1)^{r}}\ ,\label{multiple_sine}
\end{align}
with $|\bomega|=\sum_{i=1}^{r}\omega_{i}$. It satisfies the following
identities \cite{kurokawa2003multiple}: 
\begin{align}
\begin{aligned}S_{r}(|\bomega|-z|\bomega) & =S_{r}(z|\bomega)^{(-1)^{r-1}}\ ,\\
S_{r}(z+\omega_{i}|\bomega) & =S_{r}(z|\bomega)S_{r-1}(z,\bomega(i))^{-1}\ ,\\
S_{r}(Nz|\bomega) & =\prod_{0\le k_{i}\le N-1}S_{r}\left(z+\frac{\boldsymbol{k}\cdot\bomega}{N}\bigg|\bomega\right)\ ,\\
N & =\prod_{0\le k_{i}\le N-1,\boldsymbol{k}\neq0}S_{r}\left(\frac{\boldsymbol{k}\cdot\bomega}{N}\bigg|\bomega\right)\ ,\\
S_{r}(cz|c\bomega) & =S_{r}(z|\bomega)\ ,\quad\text{for}\quad c>0\ .
\end{aligned}
\label{MS_formula}
\end{align}
where $\bomega(i)=(\omega_{1},\cdots,\omega_{i-1},\omega_{i+1},\cdots,\omega_{r})$.

The formula (\ref{ZetaFormula}) yields an additional identity 
\begin{align}
S_{r}\left(z\bigg|\omega_{1},\cdots,\omega_{r-1},\frac{\omega_{r}}{N}\right)=\prod_{k=0}^{N-1}S_{r}\left(z+\frac{k\omega_{r}}{N}\bigg|\bomega\right)\ .\label{Replica_Identity}
\end{align}
This is the generalization of the identity for the hyperbolic gamma
function found in \cite{Nishioka:2013haa}.

\paragraph{The $\Upsilon$ function}

The double gamma function is used to define the $\Upsilon$ function
\cite{Zamolodchikov:1995aa,Hama:2012bg} 
\begin{align}
\Upsilon(z|\omega_{1},\omega_{2}):=\Gamma_{2}^{2}\left(\frac{|\bomega|}{2}\bigg|\omega_{1},\omega_{2}\right)\left(\Gamma_{2}(z|\omega_{1},\omega_{2})\Gamma_{2}(|\bomega|-z|\omega_{1},\omega_{2})\right)^{-1}\ .\label{Def_Upsilon}
\end{align}
satisfies several identities 
\begin{align}
\begin{aligned}\Upsilon(z+\omega_{1}|\omega_{1},\omega_{2}) & =\omega_{2}^{\frac{2z}{\omega_{2}}-1}\gamma(z/\omega_{2})\Upsilon(z|\omega_{1},\omega_{2})\ ,\\
\Upsilon(z+\omega_{2}|\omega_{1},\omega_{2}) & =\omega_{1}^{\frac{2z}{\omega_{1}}-1}\gamma(z/\omega_{1})\Upsilon(z|\omega_{1},\omega_{2})\ ,
\end{aligned}
\end{align}
where $\gamma(z):=\Gamma(z)/\Gamma(1-z)$ and the scaling law 
\begin{align}
\Upsilon(cz|c\omega_{1},c\omega_{2})=c^{\frac{(|\bomega|-2z)^{2}}{4\omega_{1}\omega_{2}}}\Upsilon(z|\omega_{1},\omega_{2})\ .
\end{align}
Some literatures including \cite{Zamolodchikov:1995aa,Hama:2012bg}
use 
\begin{align}
\Upsilon_{b}(z):=\Upsilon(z|b,1/b)\ .
\end{align}
which is sometimes denoted $\Upsilon(z)$ without the subscript.

The formula (\ref{ZetaFormula}) yields 
\begin{align}
\Upsilon\left(z\bigg|\omega_{1},\frac{\omega_{2}}{N}\right)=\prod_{k=0}^{N-1}\Upsilon\left(z+\frac{k\omega_{2}}{N}\bigg|\omega_{1},\omega_{2}\right)\ .\label{4dSurfaceFormula}
\end{align}

\section{\label{sec:The-instanton-partition}The instanton partition function}

Nekrasov's instanton partition function, \cite{Nekrasov:2002qd,Nekrasov:2003rj},
is the equivariant volume of the instanton moduli space with respect
to the action of 
\begin{equation}
U(1)_{a}\times U(1)_{\epsilon_{1}}\times U(1)_{\epsilon_{2}}\ .\label{eq:equivariant_action}
\end{equation}
The three factors correspond to (constant) gauge transformations and
to rotations in two orthogonal two-planes inside $\mathbb{R}^{4}$,
respectively. The $q$-deformed version of the partition function
counts instantons extended along an additional $\BS^1$ factor in
the geometry of circumference $\beta$. The undeformed partition function
can be recovered by letting the size of this $\BS^1$ shrink to $0$.
Our expressions for the instanton partition function are taken from
\cite{Awata:2008ed}. We use a 5$d$ parameter $\beta$ which can be
used to take the 4$d$ limit, and our conventions differ from those in
\cite{Awata:2008ed} by the substitutions 
\begin{equation}
\epsilon_{1}~\rightarrow~ i\beta\,\epsilon_{1}\ ,\qquad\epsilon_{2}~\rightarrow~ i\beta\,\epsilon_{2}\ ,\qquad a ~\rightarrow~ i\beta\, a\ .\label{eq:Awata_redefinition}
\end{equation}

The $q$-deformed version of the instanton partition function for
$G=U(N)$ and in the presence of hypermultiplets can be
expressed as follows \cite{Awata:2008ed,Sulkowski:2009ne}
\begin{align}
Z_{\text{inst}}\left(\mathfrak{q},\vec{a},\vec{m_{f}},\epsilon_{1},\epsilon_{2},\beta\right) & =\sum_{\vec{{\bf Y}}}\mathfrak{q}{}^{\left|\vec{{\bf Y}}\right|}Z_{\vec{{\bf Y}},\kappa}^{\text{5$d$-CS}}\left(\vec{a},\epsilon_{1},\epsilon_{2},\beta\right)Z_{\vec{{\bf Y}}}\left(\vec{a},\vec{m_{f}},\epsilon_{1},\epsilon_{2},\beta\right)\ ,\label{eq:instanton_partition_function}\\
Z_{\vec{{\bf Y}}}\left(\vec{a},\vec{m_{f}},\epsilon_{1},\epsilon_{2},\beta\right) & =Z_{\vec{{\bf Y}}}^{\text{vector}}\left(\vec{a},\epsilon_{1},\epsilon_{2},\beta\right)\prod_{l=1}^{N_{f}}Z_{\vec{{\bf Y}}}^{\text{hyper}}\left(\vec{a},\left(m_{f}\right)_{l},\epsilon_{1},\epsilon_{2},\beta\right)\ ,\nonumber \\
Z_{\vec{{\bf Y}}}^{\text{vector}}\left(\vec{a},\epsilon_{1},\epsilon_{2},\beta\right) & =\prod_{\left(l,i\right)\ne\left(m,j\right)}\frac{\left(Q_{l}\,Q_{m}^{-1}\,q^{Y_{li}-Y_{mj}}\,t^{j-i};q\right)_{\infty}}{\left(Q_{l}\,Q_{m}^{-1}\,q^{Y_{li}-Y_{mj}}\,t^{j-i+1};q\right)_{\infty}}\frac{\left(Q_{l}\,Q_{m}^{-1}\,t^{j-i+1};q\right)_{\infty}}{\left(Q_{l}\,Q_{m}^{-1}\,t^{j-i};q\right)_{\infty}}\ ,\label{eq:instanton_vector}\\
Z_{\vec{{\bf Y}}}^{\text{adjoint hyper}}\left(\vec{a},m_{f},\epsilon_{1},\epsilon_{2},\beta\right) & =\prod_{\left(l,i\right),\left(m,j\right)}\frac{\left(Q_{m_{f}}\,Q_{l}\,Q_{m}^{-1}\,q^{Y_{li}-Y_{mj}}\,t^{j-i+1};q\right)_{\infty}}{\left(Q_{m_{f}}^{-1}\,Q_{l}\,Q_{m}^{-1}\,q^{Y_{li}-Y_{mj}}\,t^{j-i};q\right)_{\infty}}\frac{\left(Q_{m_{f}}^{-1}\,Q_{l}\,Q_{m}^{-1}\,t^{j-i};q\right)_{\infty}}{\left(Q_{m_{f}}\,Q_{l}\,Q_{m}^{-1}\,t^{j-i+1};q\right)_{\infty}}\ ,\\
Z_{\vec{{\bf Y}}}^{\text{fund hyper}}\left(\vec{a},m_{f},\epsilon_{1},\epsilon_{2},\beta\right) & =\prod_{l=1}^{N}\prod_{j=1}^{\infty}\frac{\left(Q_{m_{f}}^{-1}\,Q_{l}\,q\,t^{-j};q\right)_{\infty}}{\left(Q_{m_{f}}^{-1}\,Q_{l}\,q^{Y_{lj}+1}\,t^{-j};q\right)_{\infty}}\ .\label{eq:instanton_hyper}
\end{align}
Another expression for the vector contribution is
\begin{align}
	\begin{aligned}
		Z_{\vec{{\bf Y}}}^{\text{vector}}\left(\vec{a},\epsilon_{1},\epsilon_{2},\beta\right) &=\prod_{l,m}^{N_{c}}\left(N_{l,m}^{\vec{{\bf Y}}}\right)^{-1}\ , \\
		N_{l,m}^{\vec{{\bf Y}}}&=\prod_{s\in Y_{l}}\left[1-e^{i\beta\left[\ell_{Y_{m}}\left(s\right)\epsilon_{1}-\left(a_{Y_{l}}\left(s\right)+1\right)\epsilon_{2}+a_{l}-a_{m}\right]}\right] \\
			&\qquad \cdot \prod_{t\in Y_{m}}\left[1-e^{i\beta\left[-\left(\ell_{Y_{l}}\left(t\right)+1\right)\epsilon_{1}+a_{Y_{m}}\left(t\right)\epsilon_{2}+a_{l}-a_{m}\right]}\right]\ .
	\end{aligned}
\end{align}

The symbols above are defined as follows:
\begin{itemize}
\item $\vec{{\bf Y}}$ is an $N$-vector of partitions ${\bf Y}_{l}$. A partition
is a non-increasing sequence of non-negative integers which stabilizes
at zero
\[
{\bf Y}_{l}=\left\{ Y_{l\,1}\ge Y_{l\,2}\ge\ldots\ge Y_{l\,n_{l}+1}=0=Y_{l\,n_{l}+2}=Y_{l\,n_{l}+3}=\ldots\right\} \ .
\]
We define
\[
\left|{\bf Y}_{l}\right|\equiv\sum_{i}Y_{li}\ ,\qquad\left\Vert {\bf Y}_{l}\right\Vert ^{2}\equiv\sum_{i}Y_{li}^{2}\ , \qquad\left|\vec{{\bf Y}}\right|\equiv\sum_{l,i}Y_{li}\ .
\]
The sum in \eqref{eq:instanton_partition_function} is over all such
partitions. A partition $\mathbf{Y}_{l}$ can be identified with a
Young diagram whose $i^{th}$ column is of height $Y_{li}$. We denote
the partition corresponding to the transposed Young diagram as ${\bf Y}_{l}^{t}$.
\item For a box $s\in Y_{l}$ with coordinates $s=\left(i,j\right)$, we
define the leg length and arm length
\[
\ell_{Y_{l}}\left(s\right)\equiv Y_{lj}^{t}-i\ ,\qquad a_{Y_{l}}\left(s\right)\equiv Y_{li}-j\ .
\]

\item $\vec{a}$ is an $N$-vector of deformation parameters corresponding to
the equivariant action of the gauge group on the instanton moduli
space. In the partition functions we compute, they are integrated
over the imaginary axis and identified with the vev of a scalar field
in the vector multiplet.
\item $\vec{m_{f}}$ is an $N_{f}$-dimensional vector of mass deformation
parameters associated to hypermultiplets. When all of the hypermultiplets
are in the fundamental representation of the gauge group, $\vec{m}_{f}$
transforms as a fundamental of the flavor symmetry group $SU(N_{f})$.
Mass deformations should be viewed as coming from a vev for a background
vector multiplet. Physical masses are the imaginary part of this vev.
\item We define 
\[
q\equiv e^{i\beta\,\epsilon_{2}}\ ,\qquad t\equiv e^{-i\beta\,\epsilon_{1}}\ ,\qquad Q_{l}\equiv e^{i\beta\, a_{l}}\ ,\qquad Q_{m_{f}}\equiv e^{i\beta\, m_{f}}\ .
\]
This definition differs from \cite{Awata:2008ed} by \eqref{eq:Awata_redefinition}.
\item The $q$-Pochhammer symbol is defined as 
\[
\left(x,q\right)_{\infty}\equiv\prod_{p=0}^{\infty}\left(1-x\, q^{p}\right)\ .
\]

\item $\mathfrak{q}$ is a classical contribution equal to\footnote{Later versions of \cite{Awata:2008ed} include a different convention
for the counting parameter, essentially redefining
\[
\mathfrak{q}\rightarrow\mathfrak{q}\,e^{-\frac{\epsilon_{1}+\epsilon_{2}}{2}}\ .
\]
There is a similar factor included in the Chern-Simons contribution.
We will not use these redefinitions.} 

\begin{itemize}
\item in 5$d$ we have 
\[
\mathfrak{q=}\,e^{-\beta \frac{8\pi^{2}}{g_{\text{YM}}^{2}}}\ .
\]

\item In a 4$d$ calculation one uses 
\[
\mathfrak{q^{\left(\text{4$d$}\right)}\equiv}\,e^{2\pi i\tau},
\]
 where $e^{2\pi i\tau}$ is minus the exponential of the one instanton
action of the conformal theory with coupling constant
\[
\tau=\frac{\theta_{\text{YM}}}{2\pi}+\frac{4\pi i}{g_{\text{YM}}^{2}}\ .
\]
If the theory is not conformal, then\footnote{We follow \cite{Tachikawa:2014dja}. $h^{\vee}\left(G\right)$ is
the dual Coxeter number and $k\left(R\right)$ is the quadratic Casimir,
normalized such that $k\left(\text{adjoint}\right)=2h^{\vee}$. For
$SU(N)$, we have $h^{\vee}=N$ and $k\left(\text{fund}\right)=1$.
The combination $2h^{\vee}\left(G\right)-k\left(R\right)$ is the
coefficient of the one-loop beta function for the 4$d$ $\mathcal{N}=2$
theory with hypermultiplets in the representation $R$.} 
\[
\mathfrak{q}^{\left(\text{4$d$}\right)}~\rightarrow~ \Lambda^{2h^{\vee}\left(G\right)-k\left(R\right)} \ ,
\]
where $\Lambda$ is the holomorphic dynamical scale. According to
\cite{Tachikawa:2014dja}, the relationship between the 5$d$ and 4$d$
partition functions is 
\[
\mathfrak{q}=\mathfrak{q}^{\left(\text{4$d$}\right)}\left(-i\beta\right)^{2h^{\vee}\left(G\right)-k\left(R\right)} \ ,
\]
and
\begin{align}
Z_{\text{inst}}^{\left(\text{4$d$}\right)}\left(\mathfrak{q}^{\left(\text{4$d$}\right)},\vec{a},\vec{m_{f}},\epsilon_{1},\epsilon_{2}\right)=\lim_{\beta\rightarrow0}Z_{\text{inst}}\left(\mathfrak{q}^{\left(\text{4d}\right)}\left(-i\beta\right)^{2h^{\vee}\left(G\right)-k\left(R\right)},\vec{a},\vec{m_{f}},\epsilon_{1},\epsilon_{2},\beta\right)\ .
\end{align}

\end{itemize}
\item In the presence of a $5d$ Chern-Simons term with parameter $\kappa$,
we have \cite{Sulkowski:2009ne} 
\[
Z_{\vec{{\bf Y}},\kappa}^{\text{5$d$-CS}}\left({\vec{a}},\epsilon_{1},\epsilon_{2},\beta\right)=\prod_{l=1}^{N}\left(Q_{l}^{\left|{\bf Y}_{l}\right|}q^{\frac{\left\Vert {\bf Y}_{l}\right\Vert ^{2}}{2}}t^{\frac{\left\Vert {\bf Y}_{l}^{t}\right\Vert ^{2}}{2}}\right)^{-\kappa}\ .
\]
An alternative version is \cite{Awata:2008ed,Tachikawa:2004ur} 
\[
Z_{\vec{{\bf Y}},\kappa}^{\text{5$d$-CS}}\left({\vec{a}},\epsilon_{1},\epsilon_{2},\beta\right)=\exp\left[i\beta\,\kappa\,\sum_{l}\sum_{\left(s,t\right)\in\mathbf{Y}_{l}}\left(a_{l}-\left(s-1\right)\epsilon_{1}-\left(t-1\right)\epsilon_{2}\right)\right]\ .
\]

\end{itemize}
The parameters $\epsilon_{1,2}$ are associated with the $\Omega$-deformation in the 4$d$ theory. They take the following values:
\begin{itemize}
\item in the derivation of the prepotential for the Seiberg-Witten solution
one takes
\[
\epsilon_{1}=-\epsilon_{2}\equiv i\hbar\ ,\qquad\beta~\rightarrow~ 0 \ ,
\]
eventually extracting the leading piece at $\hbar\rightarrow0$. 
\item in the computation on the squashed four-sphere, with squashing parameters
$\omega_{1}=\ell^{-1},\,\omega_{2}=\tilde{\ell}^{-1}$, the instanton
contribution from the north pole involves 
\[
\epsilon_{1}=\omega_{1}\ ,\qquad\epsilon_{2}=\omega_{2}\ ,
\]
and for anti-instantons from the south pole
\[
\epsilon_{1}=\omega_{2}\ ,\qquad\epsilon_{2}=\omega_{1}\ ,
\]
and we take the limit $\beta\rightarrow0$. For the squashed $\mathbb{S}^{4}$
corresponding to the $n^{th}$ supersymmetric R{\'e}nyi
entropy, we choose
\[
\omega_{1}=1\ ,\qquad\omega_{2}=\frac{1}{n}\ .
\]

\item in the computation on the five-sphere, $\beta$ is the circumference
of the circle fiber and there are three fixed points, the values at
which are given in table \ref{tab:5d_deformation_parameters}. These
correspond to the values given in \cite{Pasquetti:2016dyl,Nieri:2013vba}.
\renewcommand{\arraystretch}{1.5}
\begin{table}[t]
\begin{centering}
\begin{tabular}{|c|c|c|c|c|c|}
\hline 
\# & $\beta$ & $\epsilon_{1}$ & $\epsilon_{2}$ & $q$ & $t$\tabularnewline
\hline 
\hline 
1 & $\frac{2\pi}{\omega_{1}}$ & $\omega_{2}$ & $\omega_{3}$ & $e^{2\pi i\frac{\omega_{3}}{\omega_{1}}}$ & $e^{-2\pi i\frac{\omega_{2}}{\omega_{1}}}$\tabularnewline
\hline 
2 & $\frac{2\pi}{\omega_{2}}$ & $\omega_{3}$ & $\omega_{1}$ & $e^{2\pi i\frac{\omega_{1}}{\omega_{2}}}$ & $e^{-2\pi i\frac{\omega_{3}}{\omega_{2}}}$\tabularnewline
\hline 
3 & $\frac{2\pi}{\omega_{3}}$ & $\omega_{1}$ & $\omega_{2}$ & $e^{2\pi i\frac{\omega_{2}}{\omega_{3}}}$ & $e^{-2\pi i\frac{\omega_{1}}{\omega_{3}}}$\tabularnewline
\hline 
\end{tabular}
\par\end{centering}

\caption{\label{tab:5d_deformation_parameters}Parameters entering the $q$-deformed
Nekrasov partition function at three fixed points on a squashed $\mathbb{S}^{5}$.}
\end{table}
For the squashed $\mathbb{S}^{5}$ corresponding to the $n^{th}$ supersymmetric
R{\'e}nyi entropy, we choose 
\[
\left(\omega_{1},\omega_{2},\omega_{3}\right)=\left(1,1,\frac{1}{n}\right)\ .
\]

\end{itemize}

\bibliographystyle{JHEP}
\bibliography{SRE_surface}

\end{document}